\documentclass[aps,twocolumn,nofootinbib,groupedaddress]{revtex4-1}

\usepackage[english]{babel}
\usepackage{graphicx}
\usepackage{amsmath,amssymb,amsbsy,amstext, amsthm, simplewick, amsfonts}
\usepackage{dcolumn}

\usepackage{hyperref}
\hypersetup{
    colorlinks=true,
    linkcolor=purple,
    filecolor=purple,      
    urlcolor=purple,
    citecolor=purple
    }
    
\usepackage{comment}
\usepackage{bm}
\usepackage{xcolor}
\usepackage{braket}
\usepackage{tensor}
\usepackage{soul}
\usepackage{tikz}
\definecolor{pyblue}{RGB}{31, 119, 180}
\definecolor{pyorange}{RGB}{255, 127, 14}
\definecolor{pygreen}{RGB}{44, 160, 44}
\definecolor{pyred}{RGB}{214, 39, 40}
\usetikzlibrary{intersections, calc, arrows.meta}
\usetikzlibrary{patterns}

\newcommand{\dd}{\mathrm{d}}

\def\mpl{M_{\rm pl}}

\usepackage{colortbl}
\definecolor{pyblue}{RGB}{31, 119, 180}
\definecolor{pyred}{RGB}{214, 39, 40}
\definecolor{pygreen}{RGB}{44, 160, 44}

\newcommand{\LP}[1]{{\color{red}#1} }

\begin{document}

\title{Freezing of the renormalized one-loop primordial scalar power spectrum}

\author{Matteo Braglia$^1$, and Lucas Pinol$^2$}
\email{$^1$matteo.braglia@cern.ch $\,, \,\, ^2$lucas.pinol@phys.ens.fr}

\affiliation{$^1$Center for Cosmology and Particle Physics, New York University, 726 Broadway, New York, NY 10003, USA \\
$^2$Laboratoire de Physique de l’École Normale Supérieure, ENS, CNRS, Université PSL,\\ Sorbonne Université, Université Paris Cité, F-75005, Paris, France}

\begin{abstract}
The predictive power of cosmic inflation hinges on the existence of a conserved quantity at very large scales, called the primordial curvature perturbation, which is therefore insensitive to the details of reheating and the physics of the hot Big Bang.
Whether the classical spacetime symmetries responsible for this conservation law can enforce it at the quantum level is an old debate.
Although this was never explicitly proven, cosmologists often assume that this freezing holds at all loop orders and that inflationary predictions can always be propagated to the radiation era, with primordial fluctuations playing the role of seeds for the large-scale structures that we observe in our universe today.
In this work, by consistently using the effective field theory of inflationary fluctuations, we explicitly prove for the first time that the renormalized one-loop power spectrum of the primordial curvature perturbation freezes exactly on scales larger than its sound horizon.
\end{abstract}

\maketitle

\paragraph*{\bf Introduction.}
The detection and characterization of Cosmic Microwave Background (CMB) anisotropies established the need for a primordial mechanism to generate a nearly scale-invariant and approximately Gaussian spectrum of density perturbations~\cite{Planck:2018jri,Tristram:2021tvh}.
The leading candidate is cosmic inflation~\cite{Guth:1980zm, Linde:1981mu, Albrecht:1982wi}, and in its simplest realization, a classical field drives a period of accelerated expansion in the early universe, while its unavoidable quantum fluctuations are stretched beyond the Hubble radius, later seeding the initial cosmological overdensities~\cite{Mukhanov:1981xt, Starobinsky:1982ee, Hawking:1982cz, Guth:1982ec, Bardeen:1983qw}. 
Thanks to the growing precision of CMB and galaxy surveys, combined with a wave of innovative theoretical techniques, cosmologists have significantly narrowed the range of viable inflationary models.

This progress hinges on the existence of a gauge-invariant quantity, the curvature perturbation $\zeta$ which
classically
freezes on scales much larger than the Hubble radius~\cite{Weinberg:2003sw}\footnote{The same property holds for the two polarizations of primordial gravitational waves, $\gamma_{+,\times}$, that we consider in our companion paper~\cite{Braglia:2025cee}.
See also~\cite{Ballesteros:2024qqx} for an explicit proof of their freezing at one-loop order.
}.
Its statistics encode the properties of the inflationary dynamics, while being insensitive to the details of the subsequent poorly understood reheating phase\footnote{These considerations apply to (effectively) single-clock models.
In multifield scenarios where a non-adiabatic pressure perturbation survives on super-horizon scales a detailed description of the reheating phase is necessary~\cite{Choi:2008et,Gonzalez:2018jax,Martin:2021frd}.
}.
The super-horizon freezing of $\zeta$ is easy to check in the linear perturbation theory, either under the Slow-Roll (SR) approximation in analytical calculations (see, e.g., Ref.~\cite{Auclair:2022yxs}) or in numerical analyses.
Subtleties arise when considering nonlinearities, but there still exists a generic proof of constancy at the fully non-perturbative though classical level~\cite{Langlois:2005qp}.
However, in order to derive concrete theoretical predictions, we need to perform perturbative quantum calculations with the celebrated in-in formalism~\cite{Maldacena:2002vr, Weinberg:2005vy}.
This introduces new challenges, chief among which is the appearance of fictitious late-time divergences in cosmological correlators.

Nonlinear interactions in the early universe have two consequences. They generate connected correlation functions beyond the two-point one, i.e. primordial non-Gaussianities, and induce corrections to the Gaussian statistics via the equivalent of loop diagrams in particle physics.
As for the former, it was shown in~\cite{Maldacena:2002vr} that the various late-time divergences eventually cancel in the primordial bispectra on super-Hubble scales, thus justifying the propagation of these non-Gaussian initial conditions to the radiation era.
About the latter, in the first attempt of one-loop calculation for inflation~\cite{Weinberg:2005vy}, Weinberg wonders whether apparent late-time divergences are an artifact of truncating the calculation at one-loop order and should be resummed in a non-perturbative framework, or whether they hint at the fact that his calculation is incomplete.
This seminal work sparked an intense debate in the cosmological community.
It was claimed in~\cite{Kahya:2010xh} that indeed the curvature perturbation develops a super-Hubble growth at one loop.
But then, it was argued in~\LP{\cite{Pimentel:2012tw}} that the Lagrangian used in the prior study was incomplete, and that the constancy of $\zeta$ is recovered after incorporating backreaction of quantum fluctuations on the inflationary spacetime.
Later, Ref.~\cite{Senatore:2012ya} proposed that $\zeta$ is conserved at all loop orders.
Although these seminal works paved the way to a rigorous investigation of loop effects during inflation, none provides an explicit calculation of the renormalized power spectrum, including its finite terms where late-time divergences may hide.
We stress that this debate is crucial: nothing but the predictivity of the inflationary theory is at stake, together with its celebrated window into physics at energy scales far beyond the reach of terrestrial experiments.

\vspace{0.2cm}
In this {\em Letter}, and in the companion paper~\cite{Braglia:2025cee} to which all technical details are deferred,
we explicitly prove once and for all the large-scale conservation of the curvature perturbation power spectrum, including one-loop corrections from leading gravitational nonlinear interactions.
To carry out this analysis, we work within the framework of the Effective Field Theory (EFT) of inflationary fluctuations~\cite{Creminelli:2006xe,Cheung:2007st}, which encompasses a large class of models beyond single-field canonical inflation.
Crucially, the EFT provides a consistent way to identify the dominant self-interactions of the Nambu-Goldstone (NG) boson $\pi$, associated with the spontaneous breaking of time diffeomorphism invariance by the inflationary background.
The NG field $\pi$ is related to $\zeta$ by a simple gauge transformation that allows us to readily obtain the power spectrum of $\zeta$.
The EFT framework also supplies the counterterms required for renormalization.
These counterterms are used (i) to cancel tadpole diagrams from cubic interactions, thereby accounting for backreaction, and (ii) to absorb time-dependent UV divergences of the bare loop contributions, identified via dimensional regularization.
Finally, the EFT description allows for a perturbative renormalization of inflation at low energies, even though gravity is a non-renormalizable theory.
The resulting renormalized power spectrum of $\zeta$ is UV-finite and suppressed by $(H/\Lambda_i)^2$, the squared ratio of the Hubble scale to the strong coupling scales associated with nonlinear gravitational interactions.
It is scale invariant and depends on an arbitrary renormalization scale $\mu$, to be fixed observationally via a renormalization condition.

\vspace{0.2cm}
While loop corrections are suppressed in the SR approximation considered in this {\em Letter}, our work represents an important step toward more general scenarios, especially those breaking scale invariance. This is of particular interest for models amplifying density fluctuations at small cosmological scales, as relevant for the formation of Primordial Black Holes (PBH)~\cite{Byrnes:2025tji} and Scalar-Induced Gravitational Waves (SIGW)~\cite{Domenech:2021ztg}.
\vskip 4pt
\paragraph*{\bf Goldstone Description.} 

At leading order in the derivative expansion, the EFT action for fluctuations compatible with the symmetries of a FLRW background reads $S=\int \dd^4 x \sqrt{-g} \, \mathcal{L}$ with~\cite{Creminelli:2006xe,Cheung:2007st}
\begin{align}
\label{eq: LO EFT action} \mathcal{L}&=\frac{\mpl^2}{2}\left[R-2\left(\dot{H} + 3H^2\right)\right]+ \mpl^2 \dot{H}  g^{00}  +  \frac{M_2^4}{2}\left(\delta g^{00}\right)^2 
\end{align}
plus independent operators starting at cubic order in fluctuations.
Interestingly, these few building blocks are enough to describe gravitational nonlinear interactions together with a generic linear propagation speed $c_s$ defined through $2 M_2^4 \equiv - \dot{H} \mpl^2(1/c_s^2-1)$.
To elucidate this, we follow a well-established method to extract the leading dynamics of the propagating scalar degree of freedom, so far hidden in spacetime metric fluctuations.

In cosmology, the FLRW background solution breaks general relativity invariance under time diffeomorphisms.
In order to restore time translations and boosts, the NG boson $\pi(t,\vec{x})$ associated with this spontaneous symmetry breaking can be spelled out.
Under this so-called Stückelberg procedure, together with a technical assumption called the {\em decoupling limit}
which amounts to assuming $\epsilon\equiv-\dot{H}/H^2\ll1$ and $H/\mpl \ll 1$,  
the EFT building blocks transform as~\cite{Cheung:2007st}
\begin{align}
f(t) \rightarrow f(t+\pi)\,, \quad 
    \delta g^{00}\rightarrow -2 \dot{\pi} - \dot{\pi}^2 +  \frac{(\partial_i \pi)^2}{a^2} \,.
\end{align}
Consistently developing the Lagrangian up to quartic order,
we find
\begin{align}
\label{eq: Lpi tot}
    \mathcal{L}_\mathrm{decoup.}^{\pi} = &  \, \frac{\epsilon H^2 \mpl^2}{c_s^2} \left(1+ \eta H \pi + \frac{\eta (\eta+\eta_2)}{2}H^2\pi^2 \right) \nonumber \\ & \, \times \left[\dot{\pi}^2 - c_s^2 \frac{(\partial_i \pi)^2}{a^2}  \right] + \mathcal{O}\left(\frac{1}{c_s^2}-1\right)\,,
\end{align}
where we used $\eta\equiv\dot{\epsilon}/(H \epsilon)$ and $\eta_2\equiv\dot{\eta}/(H \eta)$.
In this expression, we have not written, and in the following we will not consider, the cubic and quartic interactions that vanish in the limit $c_s^2 \rightarrow 1$.
Our motivation is to focus on the unavoidable gravitational interactions that must be present in \textit{any} inflationary theory\footnote{The cubic part of these interactions indeed corresponds to the ``gravitational floor'' resulting in Maldacena's consistency relation~\cite{Maldacena:2002vr} that the squeezed limit bispectrum is bounded below by $1-n_s$, i.e. $\eta$ when $\epsilon\ll 1$.}, while retaining the $c_s$-dependence to remain as generic as possible.
This Lagrangian can be used to study all one-loop corrections to the power spectrum of $\pi$, and therefore of $\zeta$ using the linear relation $\zeta \approx - H \pi$ valid on large scales and in the decoupling limit~\cite{Cheung:2007st}.

Beyond free propagation, we are in fact interested in the interaction Hamiltonian, which defines the vertices in the perturbative quantum in-in calculations.
The cubic Hamiltonian is simply given by
\begin{align}
     a\mathcal{H}^{(3)}_\mathrm{int} = -a^4 \mathcal{L}^{(3)}&=- a^4\epsilon \eta H^3 \mpl^2 
    \pi \left[
    \frac{1}{c_s^2}
    \dot{\pi}^{ 2}-  \frac{(\partial_i \pi)^2}{a^2}\right]\notag\\
\label{eq:H3_pi}&\equiv\vcenter{\hbox{\begin{tikzpicture}[line width=1. pt, scale=2]
    \draw[pyred] (-0.2, 0) -- (0.0, 0);
    \draw[pyred] (0.0, 0.) -- (0.1, 0.173) ;
    \draw[pyred]  (0.0, 0.)  -- (0.1, -0.173);
    \node[draw, circle, fill=black, inner sep=1.5pt]  at (0.0,-
    0.0) {};  
\end{tikzpicture}}}
\quad+\quad
    \vcenter{\hbox{\begin{tikzpicture}[line width=1. pt, scale=2]
    \draw[pyred] (-0.2, 0) -- (0.0, 0);
    \draw[pyred] (0.0, 0.) -- (0.1, 0.173) ;
    \draw[pyred]  (0.0, 0.)  -- (0.1, -0.173);
    \node[draw, circle, fill=white, inner sep=1.5pt]  at (0.0,-
    0.0) {};  
\end{tikzpicture}}}\,,
\end{align}
where, here and in the following, we represent free propagators of $\pi$ with a red line and time-(spatial-)derivative vertices with a black (white) dot.
The relation between the conjugate momentum $p_\pi$ and $\dot{\pi}$ being nonlinear, the quartic interaction Hamiltonian is not just  $-a^3\mathcal{L}^{(4)}$~\cite{Wang:2013zva}. 
A careful Legendre transform actually leads to:
\begin{align}
    a \mathcal{H}^{(4)}_\mathrm{int} &= \frac{a^4}{2}   \epsilon \eta  H^4 \mpl^2  
   \pi^2 \left[ \frac{\eta-\eta_2}{c_s^2} 
    \dot{\pi}^2
    + (\eta+\eta_2)\frac{(\partial_i \pi)^2}{a^2}
    \right]\,\notag\\
\label{eq:H4_pi}
    &\equiv\vcenter{\hbox{\begin{tikzpicture}[line width=1. pt, scale=2]
\draw[pyred] (-0.2, 0.21) -- (0.0, 0);
\draw[pyred] (-0.2, -0.21) -- (0.0, 0);
\draw[pyred] (0.0, 0.) -- (0.2, 0.21) ;
\draw[pyred]  (0.0, 0.)  -- (0.2, -0.21);
    \node[draw, circle, fill=black, inner sep=1.5pt]  at (0.0,0.0) {};     
\end{tikzpicture}}}
\quad+\quad
    \vcenter{\hbox{\begin{tikzpicture}[line width=1. pt, scale=2]
\draw[pyred] (-0.2, 0.21) -- (0.0, 0);
\draw[pyred] (-0.2, -0.21) -- (0.0, 0);
\draw[pyred] (0.0, 0.) -- (0.2, 0.21) ;
\draw[pyred]  (0.0, 0.)  -- (0.2, -0.21);
    \node[draw, circle, fill=white, inner sep=1.5pt]  at (0.0,0.0) {};  
\end{tikzpicture}}}.
\end{align}

\vskip 4pt
\paragraph*{\bf Dimensional regularization.} As customary in QFT, the in-in integrals to compute the one-loop correction to the power spectrum of $\pi$ diverge in the UV.
We regulate such divergences using dimensional regularization~\cite{tHooft:1973mfk}, promoting the number of spatial dimensions to $d=3+\delta$. 
With this method, interactions are analytically continued to $d$ spatial dimensions through 
$
\int\frac{\dd^3k}{(2\pi)^3}\int \dd\tau\,a(\tau)\mapsto
\mu^\delta\int\frac{\dd^{3+\delta}k}{(2\pi)^{3+\delta}}\int \dd\tau\,a^{1+\delta}(\tau),
$
where $\mu$ is a renormalization scale introduced to maintain a dimensionless action, and the extra power of $\delta$ in the integration measure makes the in-in integrals convergent.
In addition, the freely propagating mode functions are also modified.
Since we will only encounter UV divergences leading to simple poles, we only need the mode functions up to linear order in $\delta$, which read~\cite{Senatore:2009cf}:
\begin{align}
\label{eq:pi_d_exp}
	\pi_k (x)\underset{\delta\to0}{=}&
    -i \frac{1}{2\sqrt{ c_s\epsilon}} \frac{ 1    }{\mpl k^{3/2}}\left(\frac{H}{ \mu c_s k}\right)^{\delta/2} ( 1-ix)e^{i 
		x} \nonumber \\
        &\,\times\Biggl(1+\frac{\delta}{2}\Biggl\{  \log \left(x\right)+ \frac{1}{1-ix} - \frac{1+ ix}{2(1- i x)} e^{-2 ix} \nonumber \\
        &\,\times\left[- i \pi +  \mathrm{Ei}(2 ix) \right]\Biggr\}\Biggr)+ \mathcal{O}(\delta^2),
\end{align}
with $x=-c_s k \tau$ and where $\mathrm{Ei}$ is the exponential integral function.

The $\delta$-correction in~\eqref{eq:pi_d_exp} makes it challenging to analytically solve the in-in loop integrals, which is part of the reason why almost no explicit loop calculation---including all finite terms at order $\delta^0$ in the final result---exists in the literature.
To solve the integrals, we will adopt a new method, 
precisely to deal with this complication.
Denoting as $\vec{p}$ the external momentum, we first introduce the dimensionless variables 
$(s,\,t)$ related to the momenta flowing in the loop, $\vec{k}\equiv v \,\lvert\vec{p}\rvert\, \hat{k}$ and $\vec{q}=\vec{p}-\vec{k}\equiv u\, \lvert\vec{p}\rvert\, \hat{q}$,  by $2 u=t + s + 1$ and $2 v=t - s + 1$. 
Second, we Taylor-expand up to linear order in $\delta\rightarrow0$ the whole integrand, keeping only the factor $t^\delta$ as a regulator of the momentum integral.
This allows us to perform easily both the $s$ and time integrals, leaving us with a unique layer of the form $\int_0^\infty \dd t\,t^\delta\,\left[F_0(t)+\delta\cdot F_1(t)\right]$, though unfortunately not analytically solvable. 
Following~\cite{Ballesteros:2024qqx}, we therefore introduce a fictitious dimensionless cutoff $t_{\rm UV}$  to separate the small and large $t$ regions.
In the region $t\in[t_{\rm UV},\,\infty[\,$, we expand the integrand in
$t\to \infty$
so the resulting integral can be performed
analytically for $\delta \neq 0$. 
In the small $t$ region, we instead regulate IR divergences with a comoving cutoff $t_{\rm IR}$, so we can safely set $\delta=0$ in $t\in[t_{\rm IR},\,t_{\rm UV}]$ and again the integral can be performed\footnote{In~\cite{Braglia:2025cee}, we explain how 
to also tackle IR divergences with dimensional regularization, following~\cite{Xue:2011hm}.}.
Then, we formally send $t_{\rm UV}\to \infty$ in both regions, and all $t_{\rm UV}$-dependent contributions cancel out. 

\begin{figure}
    \centering    \includegraphics{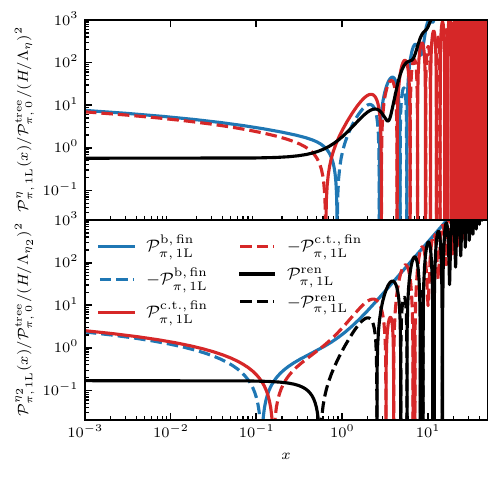}
    \vspace{-0.5cm}
    \caption{Time evolution of the finite parts of the 1-Loop power spectrum. 
            On the top and bottom panels we plot the contributions proportional to $(H/\Lambda_\eta)^2$ and $(H/\Lambda_{\eta_2})^2$.
            The counterterm finite part is defined as $   \mathcal{P}_{\pi,\,1{\rm L }}^{\rm c.t.,\,fin}=\mathcal{P}_{\pi,\,1{\rm L }}^{\mathrm{c.t.},(\Lambda,c)}+\sum_i \delta_i \mathcal{P}_{\pi,\,1{\rm L }}^{\delta_i,\, {\rm fin}}$.}
    \label{fig:P_pi}
\end{figure}

\vskip 4pt
\paragraph*{\bf Bare power spectrum.} With these premises, we can now quote the bare one-loop power spectrum:
\begin{align}
\label{eq: bare loop PS full}
\mathcal{P}_{\pi,1{\rm L }}^{\rm bare}(x)=&\,
\vcenter{\hbox{\begin{tikzpicture}[line width=1. pt, scale=1.5]
    \node[draw, circle, fill, inner sep=1.2pt] (v1) at (-0.2,0.2) {};    
    \node[draw, circle, fill, inner sep=1.2pt] (v2) at (0.2,0.2) {};
    \draw[pyred] (-0.35-0.2,0.2) -- (v1);
    \draw[pyred] (0.35+0.2,0.2) -- (v2);
    \draw[thick, pyred] (0.0,0.0) arc[start angle=270, end angle=90, radius=0.2];
    \draw[thick, pyred](0.0,0.0) arc[start angle=-90, end angle=90, radius=0.2];
    \node[draw, circle, fill, inner sep=1.2pt] at (-0.2,0.2) {};    
    \node[draw, circle, fill, inner sep=1.2pt]  at (0.2,0.2) {};
\end{tikzpicture}}}
+
\vcenter{\hbox{\begin{tikzpicture}[line width=1. pt, scale=1.5]
    \node[draw, circle, fill, inner sep=1.2pt] (v1) at (-0.2,0.2) {};    
    \node[draw, circle, fill, inner sep=1.2pt] (v2) at (0.2,0.2) {};
    \draw[pyred] (-0.35-0.2,0.2) -- (v1);
    \draw[pyred] (0.35+0.2,0.2) -- (v2);
    \draw[thick, pyred] (0.0,0.0) arc[start angle=270, end angle=90, radius=0.2];
    \draw[thick, pyred](0.0,0.0) arc[start angle=-90, end angle=90, radius=0.2];
    \node[draw, circle, fill=white,inner sep=1.2pt] at (-0.2,0.2) {};    
    \node[draw, circle, fill=white, inner sep=1.2pt]  at (0.2,0.2) {};
\end{tikzpicture}}}
+
\vcenter{\hbox{\begin{tikzpicture}[line width=1. pt, scale=1.5]
    \node[draw, circle, fill, inner sep=1.2pt] (v1) at (-0.2,0.2) {};    
    \node[draw, circle, fill, inner sep=1.2pt] (v2) at (0.2,0.2) {};
    \draw[pyred] (-0.35-0.2,0.2) -- (v1);
    \draw[pyred] (0.35+0.2,0.2) -- (v2);
    \draw[thick, pyred] (0.0,0.0) arc[start angle=270, end angle=90, radius=0.2];
    \draw[thick, pyred](0.0,0.0) arc[start angle=-90, end angle=90, radius=0.2];
    \node[draw, circle, fill=white, inner sep=1.2pt] at (-0.2,0.2) {};    
    \node[draw, circle, fill, inner sep=1.2pt]  at (0.2,0.2) {};
\end{tikzpicture}}} \notag \\
& \, +
\vcenter{\hbox{\begin{tikzpicture}[line width=1. pt, scale=1.5]
    \node[draw, circle, fill, inner sep=1.2pt] (v1) at (-0.2,0.2) {};    
    \node[draw, circle, fill, inner sep=1.2pt] (v2) at (0.2,0.2) {};
    \draw[pyred] (-0.35-0.2,0.2) -- (v1);
    \draw[pyred] (0.35+0.2,0.2) -- (v2);
    \draw[thick, pyred] (0.0,0.0) arc[start angle=270, end angle=90, radius=0.2];
    \draw[thick, pyred](0.0,0.0) arc[start angle=-90, end angle=90, radius=0.2];
    \node[draw, circle, fill, inner sep=1.2pt] at (-0.2,0.2) {};    
    \node[draw, circle, fill=white, inner sep=1.2pt]  at (0.2,0.2) {};
\end{tikzpicture}}}
+
\vcenter{\hbox{\begin{tikzpicture}[line width=1. pt, scale=1.5]
    \node[draw, circle, fill, inner sep=1.2pt] (v) at (0,0) {};
    \draw[pyred] (-0.35,0) -- (v);
    \draw[pyred] (0.35,0) -- (v);
    \draw[thick, pyred] (0.0,0.0) arc[start angle=270, end angle=90, radius=0.2];
    \draw[thick, pyred](0.0,0.0) arc[start angle=-90, end angle=90, radius=0.2];
    \node[draw, circle, fill, inner sep=1.2pt] at (0,0) {};
\end{tikzpicture}}}
+
\vcenter{\hbox{\begin{tikzpicture}[line width=1. pt, scale=1.5]
    \node[draw, circle, fill, inner sep=1.2pt] (v) at (0,0) {};
    \draw[pyred] (-0.35,0) -- (v);
    \draw[pyred] (0.35,0) -- (v);
    \draw[thick, pyred] (0.0,0.0) arc[start angle=270, end angle=90, radius=0.2];
    \draw[thick, pyred](0.0,0.0) arc[start angle=-90, end angle=90, radius=0.2];
    \node[draw, circle, fill=white, inner sep=1.2pt]  at (0,0) {};
\end{tikzpicture}}}
\end{align}
\begin{align}=&  -(1+x^2)\mathcal{P}_{\pi,\,0}^{\rm tree^2}H^2 \Biggl[\frac{\eta(\eta-2\eta_2)}{4 }\frac{1}{\delta} \notag  \\ & \,+ \frac{\eta(\eta-\eta_2)}{2}\log t_\mathrm{IR} \Biggr] + \mathcal{P}_{\pi,1{\rm L }}^{\rm b,\rm fin}(x) \notag \\
\underset{x\rightarrow 0}{\sim}& -\mathcal{P}_{\pi,\,0}^{\rm tree^2}H^2\frac{1}{2}\eta (2\eta-\eta_2)\log x \notag \,,  
\end{align}
where in the last line we show explicitly the dangerous late-time logarithmic divergence, from the 
finite terms $\mathcal{P}_{\pi,1{\rm L }}^{\rm b,\rm fin}(x)$, and whose exact time evolution is shown in Fig.~\ref{fig:P_pi} in blue. 
Time-dependent UV divergences show up as simple poles in $\delta$, and the power spectrum contains IR logarithmic divergences too.
We also defined the late-time, tree-level, dimensionless power spectrum of $\pi$ as $\mathcal{P}_{\pi,\,0}^{\rm tree}\equiv  \underset{x\rightarrow 0}{\mathrm{lim}} \,\, p^3 \lvert\pi_p(x)\rvert^2/2\pi^2=\left(8 \pi^2 \epsilon\,  c_s \mpl^2\right)^{-1}$.

\vskip 4pt
\paragraph*{\bf Cancellation of all non-1PI diagrams.} Because of the cubic interactions, tadpole diagrams are not vanishing any more at the one-loop level.
This is a problem as we want to use a perturbation theory defined with zero mean fluctuations, i.e. $\braket{\pi}=0$.
In order to enforce this property, we consider all one-point counterterms emerging from unitary gauge operators compatible with the EFT symmetries,
\begin{equation}
\label{eq:L_ct_1}
    \mathcal{L}_\mathrm{c.t.} \supset -\mpl^2 \delta\Lambda(t) - \delta c(t) \, \delta g^{00}\, 
    \end{equation}
    from which we can identify the following linear interaction Hamiltonian,
    \begin{equation}
    a\mathcal{H}_{\rm int}^{(1)}= a^4\mpl^2 \delta\dot{\Lambda}\, \pi  + 2 a^4\delta c \, \dot{\pi}\,.
\end{equation}
With an appropriate choice of the time-dependent coefficients $\delta\dot{\Lambda}(t),\, \delta c(t)$, tadpole and counterterm contributions to the 1-point function of $\pi$ can cancel each other:
\begin{equation}
\label{eq:tadpole_cancellation}
\langle\pi_{\vec{p}}(\tau)\rangle'=
\vcenter{\hbox{\begin{tikzpicture}[line width=1. pt, scale=2]
    \draw[pyred] (0,-0.2) -- (0,0);
    \draw[thick, pyred] (0.0,0.0) arc[start angle=270, end angle=90, radius=0.15];
    \draw[thick, pyred](0.0,0.0) arc[start angle=-90, end angle=90, radius=0.15];
    \node[draw, circle, fill, inner sep=1.5pt]  at (0.0,0.0) {};    
\end{tikzpicture}}}
+
\vcenter{\hbox{\begin{tikzpicture}[line width=1. pt, scale=2]
    \draw[pyred] (0,-0.2) -- (0,0);
    \draw[thick, pyred] (0.0,0.0) arc[start angle=270, end angle=90, radius=0.15];
    \draw[thick, pyred](0.0,0.0) arc[start angle=-90, end angle=90, radius=0.15];
    \node[draw, circle, fill=white, inner sep=1.5pt]  at (0.0,0.0) {};    
\end{tikzpicture}}}
+
\vcenter{\hbox{\begin{tikzpicture}[line width=1. pt, scale=2]
    \draw[pyred] (0,-0.3) -- (0,-0.06);

\node[draw,  minimum size=7pt, inner sep=0pt,
          path picture={\draw[line width=1pt] 
            (path picture bounding box.south west) -- (path picture bounding box.north east)
            (path picture bounding box.north west) -- (path picture bounding box.south east);}
         ] (X) at (0,0) {};  
         
\node at (0, 0.2) {$\delta\dot{\Lambda}$};
\end{tikzpicture}}}+
\vcenter{\hbox{\begin{tikzpicture}[line width=1. pt, scale=2]
    \draw[pyred] (0,-0.3) -- (0,-0.06);

\node[draw,  minimum size=7pt, inner sep=0pt,
          path picture={\draw[line width=1pt] 
            (path picture bounding box.south west) -- (path picture bounding box.north east)
            (path picture bounding box.north west) -- (path picture bounding box.south east);}
         ] (X) at (0,0) {};  
         
\node at (0, 0.2) {$\delta c$};
\end{tikzpicture}}}=0\,.
\end{equation}
We can actually be more ambitious and ask that the \textit{effective one-point interactions} appearing in all non-one-particle-irreducible (non-1PI) diagrams of the theory be exactly vanishing, which sets independently
    \begin{equation}
\label{eq: solutions tadpole counterterms}
       \delta{\dot{\Lambda}}=\frac{3}{16\pi^2 c_s^3}\frac{H^5}{\mpl^2}\eta\,, \quad  \delta{c}=-\frac{1}{16\pi^2 c_s^3}H^4\eta\,.
\end{equation}

\vskip 4pt
\paragraph*{\bf Renormalization.} 
We now need to
consider all quadratic counterterms 
in the interaction Hamiltonian that can contribute to the two-point function, as required to perform the renormalization of the quadratic theory.
To obtain them, we start from the most generic counterterm Lagrangian compatible with EFT symmetries:
\begin{align}
    \mathcal{L}_\mathrm{c.t.} =& -\mpl^2 \delta\Lambda - \delta c \,\delta g^{00}  +  \frac{\delta M_2^4}{2}\left( \delta g^{00}\right)^2 \notag\\&\label{eq:L_ct_2} - \frac{\bar{M}_3^2}{2} \delta K^2 + \frac{m_3^2}{2} \left(\partial_i \delta g^{00}\right)^2 \,.
\end{align}
In the first line, we kept the operators in~\eqref{eq:L_ct_1}, as they also contribute quadratic interactions due to the non-linearly realized symmetries of the EFT, and also considered a possible sound speed renormalization through $M_2^4 \rightarrow M_2^4 +  \delta M_2^4$.
In the second line, we added operators at the next order in the derivative expansion; as we are going to see, those are instrumental in absorbing the time-dependent UV divergences from bare loop contributions~\eqref{eq: bare loop PS full}.

After restoring $\pi$, the second and third operators in~~\eqref{eq:L_ct_2} contain $\dot{\pi}$'s  and therefore, via the Legendre transform of the full Lagrangian, result in non-trivial modifications to the  quadratic Hamiltonian:
\begin{align}
    \mathcal{H}^{(2)}(\pi,p_\pi)\supset&-\frac{p_\pi^2}{4 a^3  \left[-\dot{H}\mpl^2+\delta c+2M_{\rm2}^4+2\,\delta M_2^4\right]}\notag\\&-a^3\left[\dot{H}\mpl^2-\delta c\right]\frac{(\partial\pi)^2}{a^2}\label{eq:Ham_np_2}.
\end{align}
While $\delta c$ was already fixed by the condition~\eqref{eq:tadpole_cancellation} enforcing tadpole cancellation, $\delta M_2^4$  is still free and only starts contributing at quadratic order.
Therefore, we expand the Hamiltonian at linear order in $\delta M_2^4$ and treat the resulting quadratic term as an interaction. 
This, together with the other operators in~\eqref{eq:L_ct_2}, allows us to write the quadratic Hamiltonian as 
the free one,
\begin{align}
    \mathcal{H}_{\rm free}(\pi,p_\pi)=\frac{c_s^2}{4 a^3 \epsilon H^2 \mpl^2}p_\pi^2-a \epsilon H^2 \mpl^2 (\partial_i \pi)^2\,,\label{eq:H_free}
\end{align}
plus quadratic interactions,
\begin{align}
\label{eq:Ham_count_mass}
    a \mathcal{H}_{\rm int}^{(2)} = &  \frac{a^4 \mpl^2}{2} \,\delta\ddot{\Lambda}\, \pi^2-2 a^4\, \left(\delta \dot{c}\,-\,\eta H\,\delta c \right)\pi\dot{\pi}\\
    & \, +a^4 \mpl^2 \left[\epsilon H^2\delta_{c_s^2} \dot{\pi}^2
    +  \delta_1 \frac{ \left(\partial^2\pi\right)^2}{a^4}+\delta_2  \frac{ \left(\partial_i\dot{\pi}\right)^2}{a^2}\right]\,. \notag
 \end{align}   
To match usual notations, we have  redefined
$\epsilon = - \frac{\dot{H}}{H^2} \mapsto \epsilon=-\frac{\dot{H}}{H^2}+\frac{\delta c}{H^2\mpl^2}$.
The effects of the backreaction of the loop corrections onto the background evolution, encoded in~\eqref{eq:tadpole_cancellation}, are thus effectively accounted for by modifying the evolution of the free theory.
In particular,~\eqref{eq: bare loop PS full} is still valid once this $\epsilon$-redefinition is considered.
   
Using~\eqref{eq: solutions tadpole counterterms}, the terms in the first line of~\eqref{eq:Ham_count_mass} give the following contribution to the power spectrum:
\begin{align}
 &  \mathcal{P}_{\pi,\,1{\rm L }}^{\mathrm{c.t.},(\Lambda,\,c)}(x) ={\hbox{\begin{tikzpicture}[line width=1. pt, scale=2]
    \draw[pyred] (-0.35,0) -- (-0.06,0);
    \draw[pyred] (0.06,0) -- (0.35,0) ;
    \node at (0, 0.2) {$\delta\ddot{\Lambda}$};
\node[draw,  minimum size=7pt, inner sep=0pt,
          path picture={\draw[line width=1pt] 
            (path picture bounding box.south west) -- (path picture bounding box.north east)
            (path picture bounding box.north west) -- (path picture bounding box.south east);}
         ] (X) at (0,0) {};  
\end{tikzpicture}}}+{\hbox{\begin{tikzpicture}[line width=1. pt, scale=2]
    \draw[pyred] (-0.35,0) -- (-0.06,0);
    \draw[pyred] (0.06,0) -- (0.35,0) ;
    \node at (0, 0.2) {$\delta\dot{c}$};
\node[draw,  minimum size=7pt, inner sep=0pt,
          path picture={\draw[line width=1pt] 
            (path picture bounding box.south west) -- (path picture bounding box.north east)
            (path picture bounding box.north west) -- (path picture bounding box.south east);}
         ] (X) at (0,0) {};  
\end{tikzpicture}}}\\=& \,\mathcal{P}_{\pi,\,0}^{\rm tree^2} H^2\frac{ \eta  (\eta_2-2 \eta ) }{4}\left[4+\left(e^{-2
   i x} (x-i)^2 {\rm Ei}(2 i 
   x) +{\rm c.c.}\right)\right]\notag
      \\ &
   \underset{x\rightarrow 0}{\sim} \mathcal{P}_{\pi,\,0}^{\rm tree^2}H^2\frac{1}{2}\eta (2\eta-\eta_2)\log x \notag \,.
\end{align}
This is a very important result. It proves that, by properly taking into account backreaction, the quantity $ \mathcal{P}^{\rm bare}_{\pi,\,1{\rm L }}(x)+\mathcal{P}_{\pi,\,1{\rm L }}^{\mathrm{c.t.},(\Lambda,\,c)}(x)$ becomes time-independent on super-Hubble scales. 
The time-independence is a \textit{built-in} feature of the system we have been considering, as we only asked consistency of the perturbation theory to cancel tadpoles; at no point did we \textit{tune} the size of the quadratic counterterms to cancel the late-time divergences.

On the other hand, the coefficients $\delta_i$ of the terms in the second line of~\eqref{eq:Ham_count_mass}, that are uniquely mapped to the Wilson functions $\delta M_2^4, \, \bar{M}_3^2,\,m_3^2$ in~\eqref{eq:L_ct_2}, have yet to be fixed. 
They give the following contributions: 
\begin{align}
\label{eq:Pzeta_counterterms}
&\mathcal{P}_{\pi,\,1{\rm L }}^{\mathrm{c.t.},(\delta_i)}(x)=
\vcenter{\hbox{\begin{tikzpicture}[line width=1. pt, scale=2]
    \draw[pyred] (-0.35,0) -- (-0.06,0);
    \draw[pyred] (0.06,0) -- (0.35,0) ;
\node[draw, circle, minimum size=7pt, inner sep=0pt,
          path picture={\draw[line width=1pt] 
            (path picture bounding box.south west) -- (path picture bounding box.north east)
            (path picture bounding box.north west) -- (path picture bounding box.south east);}
         ] (X) at (0,0) {};       
\node at (0, 0.2) {$\delta_{c_s^2}$};
\end{tikzpicture}}}+\vcenter{\hbox{\begin{tikzpicture}[line width=1. pt, scale=2]
    \draw[pyred] (-0.35,0) -- (-0.06,0);
    \draw[pyred] (0.06,0) -- (0.35,0) ;
\node[draw, circle, minimum size=7pt, inner sep=0pt,
          path picture={\draw[line width=1pt] 
            (path picture bounding box.south west) -- (path picture bounding box.north east)
            (path picture bounding box.north west) -- (path picture bounding box.south east);}
         ] (X) at (0,0) {};       
\node at (0, 0.2) {$\delta_1$};
\end{tikzpicture}}}+\vcenter{\hbox{\begin{tikzpicture}[line width=1. pt, scale=2]
    \draw[pyred] (-0.35,0) -- (-0.06,0);
    \draw[pyred] (0.06,0) -- (0.35,0) ;
\node[draw, circle, minimum size=7pt, inner sep=0pt,
          path picture={\draw[line width=1pt] 
            (path picture bounding box.south west) -- (path picture bounding box.north east)
            (path picture bounding box.north west) -- (path picture bounding box.south east);}
         ] (X) at (0,0) {};       
\node at (0, 0.2) {$\delta_2$};
\end{tikzpicture}}}
\\&=\delta_{c_s^2}\,\mathcal{P}_{\pi,\,0}^{\rm tree^2}\,\pi ^2 c_s^3  \mpl^2 \epsilon 
   \left(-1+x^2\right)\left[1 +\delta \log\left(\frac{ H}{\mu}\right) \right]\notag\\
   &-\delta_1\,\mathcal{P}_{\pi,\,0}^{\rm tree^2}\,    \frac{\pi ^2 \mpl^2   }{2 c_s }  \left(5+5  x^2+2
    x^4\right) \left[1 +\delta \log\left(\frac{ H}{\mu}\right) \right]\notag\\
   &+\delta_2\,\mathcal{P}_{\pi,\,0}^{\rm tree^2}\, \frac{\pi ^2 c_s \mpl^2
   }{2 } \left(1+x^2+2 x^4\right)    \left[1 +\delta \log\left(\frac{ H}{\mu}\right) \right]\notag\\&+\delta_{c_s^2} \mathcal{P}_{\pi,\,1{\rm L }}^{{\delta_{c_s^2}},\,{\rm fin}}(x)+ \delta_{1} \mathcal{P}_{\pi,\,1{\rm L }}^{{\delta_1},\,{\rm fin}}(x)+\delta_{2} \mathcal{P}_{\pi,\,1{\rm L }}^{{\delta_2},\,{\rm fin}}(x) \notag \,,
\end{align}
where the functions $\mathcal{P}_{\pi,\,1{\rm L }}^{\delta_{i},\,{\rm fin}}(x) \sim \mathcal{O(\delta)}$ are finite at $x=0$. Requiring that the apparent order $\mathcal{O}(\delta^0)$ parts of~\eqref{eq:Pzeta_counterterms} cancel the UV divergences in~\eqref{eq: bare loop PS full} for all times, we get 
\begin{align}
    \delta_1=& - \frac{1}{\delta} \frac{H^2}{\mpl^2}\frac{c_s \eta   ( \eta -2
   \eta_2)}{8 \pi ^2 
   }-\frac{c_s^4\epsilon}{2}\,\delta_{c_s^2},\notag\\\delta_2=&  -\frac{1}{\delta} \frac{H^2}{\mpl^2}\frac{\eta   ( \eta -2
   \eta_2)}{8 \pi ^2 c_s
   }+c_s^2\epsilon\,\delta_{c_s^2},
\end{align}
where $\delta_{c_s^2}$ is arbitrary. We therefore simply set $\delta_{c_s^2}=0$, which agrees with the EFT power counting: gravitational interactions have dimensions strictly greater than four and, as such, modify the dispersion relation at high energies, but do not  radiatively affect the linear propagation of the free theory.
In other words: the sound speed does not get renormalized.

Summing all diagrams,
we define the renormalized, dimensionless, scale-invariant, one-loop power spectrum of  $\pi(t,\vec{x})$ as $\mathcal{P}_{\pi,\mathrm{1L}}^{\mathrm{ren}}(x)=\mathcal{P}_{\pi,1{\rm L }}^{\rm bare}(x)+\mathcal{P}_{\pi,\,1{\rm L }}^{\mathrm{c.t.},(\Lambda,\,c)}(x)+\mathcal{P}_{\pi,\,1{\rm L }}^{\mathrm{c.t.},(\delta_i)}(x)$, and show its time evolution in Fig.~\ref{fig:P_pi}. Quickly after sound horizon crossing at $x\approx1$, the renormalized power spectrum {\em freezes} to its late time value:
\begin{widetext}
    \begin{equation}
        \frac{\mathcal{P}_{\pi,\mathrm{1L},0}^{\mathrm{ren}}}{\mathcal{P}_{\pi,0}^{\mathrm{tree}}}  =  \, \frac{1}{8\pi^2} \left(\frac{H}{\Lambda_\eta}\right)^2 \left[\frac{269-290\log(2)}{160} - \frac{1}{4}\log\left(\frac{H}{\mu}\sqrt{\frac{\pi}{4 c_s^2 e^{\gamma_E}}}t_{\rm IR}^2\right)\right]  + \frac{1}{8\pi^2}\left(\frac{H}{\Lambda_{\eta_2}}\right)^2 \left[\frac{11}{24} + \frac{1}{2}\log\left(\frac{H}{\mu}\sqrt{\frac{\pi}{4 c_s^2 e^{\gamma_E}}}t_{\rm IR}\right)\right] \,,
        \label{eq:P_ren_late_time}
    \end{equation}
    \end{widetext}
    where we have defined the strong coupling scales $\Lambda^2_\eta \equiv \mpl^2  \epsilon  c_s/\eta^2$ and $\Lambda^2 _{\eta_2} \equiv \mpl^2  \epsilon  c_s/(\eta \eta_2)$.

The one-loop renormalized power spectrum of the curvature perturbation at the end of inflation is simply obtained using the linear relation $\zeta\approx-H\pi$ valid on super-Hubble scales and in the decoupling limit, so that $\mathcal{P}_{\zeta,0}^{\mathrm{ren}}(\mu)=H^2\left(\mathcal{P}_{\pi,0}^{\mathrm{tree}}+\mathcal{P}_{\pi,\mathrm{1L},0}^{\mathrm{ren}}+\cdots\right)$. A simple renormalization condition in our scale-invariant scenario is to fix $\mathcal{P}_{\zeta,0}^{\mathrm{ren}}(\mu=H)=A_s\simeq 2.1\times10^{-9}$~\cite{Planck:2018jri}.
As a result, we can always rewrite $\mathcal{P}_{\zeta,0}^{\mathrm{ren}}(\mu)$ in terms of the observed value of $A_s$, and a correction suppressed by two inverse powers of the high energy EFT scales and proportional to the logarithmic running $\mathrm{log}(H/\mu)$.
Unfortunately, this running is not observable in scale-invariant scenarios, as there is no scale to run with.
Despite this unobservability, our results can still be used to place perturbativity bounds by requiring $\mathcal{P}_{\pi,\mathrm{1L},0}^{\mathrm{ren}}\ll\mathcal{P}_{\pi,0}^{\mathrm{tree}}$.
This sets bounds on the strong coupling scales, that are easily seen to hold at CMB scales, confirming that loop corrections from gravitational nonlinearities are indeed negligible~\cite{Braglia:2025cee}.

The final result displayed in Eq.~\eqref{eq:P_ren_late_time} still contains an explicit dependence on the IR cutoff, appearing as $\log \left(p/\Lambda_\mathrm{IR} \right)$ once comoving scales are restored. While several compelling works have argued that this dependence is spurious and should vanish when predictions are expressed in terms of primordial fluctuations defined in a coordinate system tied to cosmological observations~\cite{Urakawa:2010it,Urakawa:2010kr,Senatore:2012nq}—see also~\cite{Tanaka:2011aj, Pajer:2013ana} for related arguments in the context of the squeezed-limit bispectrum—the status of IR divergences remains debated. For a review, see~\cite{Seery:2010kh}, and for a recent counterpoint advocating a physical (rather than comoving) IR cutoff, see~\cite{Huenupi:2024ksc}.
In this {\em Letter}, we have decided to keep track of all IR divergences in the one-loop power spectrum of $\pi$, but we postpone the burden of propagating their (potentially null) effects to cosmological observables to a future work.

\vskip 4pt
\paragraph*{\bf Conclusions.} In this {\em Letter}, we explicitly computed all contributions to the renormalized one-loop primordial scalar power spectrum, and demonstrated that it freezes shortly after sound horizon crossing.
While the loop-order constancy of $\zeta$
has been claimed in several previous works, we provide the first 
full calculation which includes (i) identifying all time-dependent UV divergences appearing as simple poles in $\delta$ in dimensional regularization; (ii) computing all finite terms at order $\delta^0$; (iii) canceling all non-1PI diagrams and time-dependent UV divergences by considering appropriate counterterms; (iv) observing the exact cancellation of late-time divergences as a result of appropriately taking into account backreaction and the non-linearly realized symmetries of the system.
This result is important: it confirms that effectively single-clock inflationary predictions are reliable and may safely be propagated as initial conditions to the radiation era without the need for a detailed description of reheating. 

But the significance of our work also lies in the methodology we developed.
This is especially timely given the recent debate over whether loop corrections may spoil perturbativity in scenarios where small-scale density perturbations are enhanced and may generate PBH or SIGW~\cite{Cheng:2021lif,Inomata:2022yte,Kristiano:2022maq,Riotto:2023hoz,Firouzjahi:2023aum,Choudhury:2023vuj,Motohashi:2023syh,Franciolini:2023lgy,Tasinato:2023ukp,Cheng:2023ikq,Fumagalli:2023hpa,Maity:2023qzw,Davies:2023hhn,Iacconi:2023ggt,Inomata:2024lud,Firouzjahi:2024psd,Caravano:2024tlp,Braglia:2024zsl,Kawaguchi:2024lsw,Ballesteros:2024zdp,Kristiano:2024vst,Kawaguchi:2024rsv,Fumagalli:2024jzz,Caravano:2024moy,Ruiz:2024weh,Firouzjahi:2024sce,Sheikhahmadi:2024peu,Inomata:2025bqw,Fang:2025vhi,Firouzjahi:2025gja,Firouzjahi:2025ihn,Inomata:2025pqa}. 
Importantly, in those scenarios, the non-linear interactions involved are precisely the gravitational ones that we have analyzed in this {\em Letter}.
Since such models typically feature strong deviations from scale invariance and violations of SR, they fall outside the regime of validity of our present results.
Nonetheless, the procedure we have established here is specifically designed to compute all contributions to the renormalized power spectrum in a transparent and explicit manner, allowing its late-time behavior to be accurately assessed.
Applying our methodology to these more general cases will be essential for obtaining theoretically consistent predictions, which may soon be tested by upcoming gravitational-wave observatories~\cite{Inomata:2018epa,LISACosmologyWorkingGroup:2023njw,LISACosmologyWorkingGroup:2024hsc,LISACosmologyWorkingGroup:2025vdz,Cecchini:2025oks}.

\vskip 4pt
\begin{acknowledgments}
We are grateful to
Guillermo Ballesteros,
Vadim Briaud, Sebastián Céspedes,
Jacopo Fumagalli, Jesús Gambín Egea,
Laura Iacconi, Riccardo Impavido,
Ryodai Kawaguchi,
Scott Melville,
Enrico Pajer,
Sébastien Renaux-Petel, 
Flavio Riccardi,
Filippo Vernizzi,
Denis Werth
and all the participants of the workshop \href{https://indico.cern.ch/event/1433472/}{Looping in the Primordial Universe} for interesting discussions related to this article. M.B. would like to thank LPENS for
hospitality during the latest stages of this project. M.B. acknowledges travel support through the India-Italy mobility
program `RELIC’ (INT/Italy/P-39/2022 (ER)).

\end{acknowledgments}

\bibliography{biblio}

\begin{thebibliography}{75}%
\makeatletter
\providecommand \@ifxundefined [1]{%
 \@ifx{#1\undefined}
}%
\providecommand \@ifnum [1]{%
 \ifnum #1\expandafter \@firstoftwo
 \else \expandafter \@secondoftwo
 \fi
}%
\providecommand \@ifx [1]{%
 \ifx #1\expandafter \@firstoftwo
 \else \expandafter \@secondoftwo
 \fi
}%
\providecommand \natexlab [1]{#1}%
\providecommand \enquote  [1]{``#1''}%
\providecommand \bibnamefont  [1]{#1}%
\providecommand \bibfnamefont [1]{#1}%
\providecommand \citenamefont [1]{#1}%
\providecommand \href@noop [0]{\@secondoftwo}%
\providecommand \href [0]{\begingroup \@sanitize@url \@href}%
\providecommand \@href[1]{\@@startlink{#1}\@@href}%
\providecommand \@@href[1]{\endgroup#1\@@endlink}%
\providecommand \@sanitize@url [0]{\catcode `\\12\catcode `\$12\catcode `\&12\catcode `\#12\catcode `\^12\catcode `\_12\catcode `\%12\relax}%
\providecommand \@@startlink[1]{}%
\providecommand \@@endlink[0]{}%
\providecommand \url  [0]{\begingroup\@sanitize@url \@url }%
\providecommand \@url [1]{\endgroup\@href {#1}{\urlprefix }}%
\providecommand \urlprefix  [0]{URL }%
\providecommand \Eprint [0]{\href }%
\providecommand \doibase [0]{http://dx.doi.org/}%
\providecommand \selectlanguage [0]{\@gobble}%
\providecommand \bibinfo  [0]{\@secondoftwo}%
\providecommand \bibfield  [0]{\@secondoftwo}%
\providecommand \translation [1]{[#1]}%
\providecommand \BibitemOpen [0]{}%
\providecommand \bibitemStop [0]{}%
\providecommand \bibitemNoStop [0]{.\EOS\space}%
\providecommand \EOS [0]{\spacefactor3000\relax}%
\providecommand \BibitemShut  [1]{\csname bibitem#1\endcsname}%
\let\auto@bib@innerbib\@empty
\bibitem [{\citenamefont {Akrami}\ \emph {et~al.}(2020)\citenamefont {Akrami} \emph {et~al.}}]{Planck:2018jri}%
  \BibitemOpen
  \bibfield  {author} {\bibinfo {author} {\bibfnamefont {Y.}~\bibnamefont {Akrami}} \emph {et~al.} (\bibinfo {collaboration} {Planck}),\ }\href {\doibase 10.1051/0004-6361/201833887} {\bibfield  {journal} {\bibinfo  {journal} {Astron. Astrophys.}\ }\textbf {\bibinfo {volume} {641}},\ \bibinfo {pages} {A10} (\bibinfo {year} {2020})},\ \Eprint {http://arxiv.org/abs/1807.06211} {arXiv:1807.06211 [astro-ph.CO]} \BibitemShut {NoStop}%
\bibitem [{\citenamefont {Tristram}\ \emph {et~al.}(2022)\citenamefont {Tristram} \emph {et~al.}}]{Tristram:2021tvh}%
  \BibitemOpen
  \bibfield  {author} {\bibinfo {author} {\bibfnamefont {M.}~\bibnamefont {Tristram}} \emph {et~al.},\ }\href {\doibase 10.1103/PhysRevD.105.083524} {\bibfield  {journal} {\bibinfo  {journal} {Phys. Rev. D}\ }\textbf {\bibinfo {volume} {105}},\ \bibinfo {pages} {083524} (\bibinfo {year} {2022})},\ \Eprint {http://arxiv.org/abs/2112.07961} {arXiv:2112.07961 [astro-ph.CO]} \BibitemShut {NoStop}%
\bibitem [{\citenamefont {Guth}(1981)}]{Guth:1980zm}%
  \BibitemOpen
  \bibfield  {author} {\bibinfo {author} {\bibfnamefont {A.~H.}\ \bibnamefont {Guth}},\ }\href {\doibase 10.1103/PhysRevD.23.347} {\bibfield  {journal} {\bibinfo  {journal} {Phys. Rev. D}\ }\textbf {\bibinfo {volume} {23}},\ \bibinfo {pages} {347} (\bibinfo {year} {1981})}\BibitemShut {NoStop}%
\bibitem [{\citenamefont {Linde}(1982)}]{Linde:1981mu}%
  \BibitemOpen
  \bibfield  {author} {\bibinfo {author} {\bibfnamefont {A.~D.}\ \bibnamefont {Linde}},\ }\href {\doibase 10.1016/0370-2693(82)91219-9} {\bibfield  {journal} {\bibinfo  {journal} {Phys. Lett. B}\ }\textbf {\bibinfo {volume} {108}},\ \bibinfo {pages} {389} (\bibinfo {year} {1982})}\BibitemShut {NoStop}%
\bibitem [{\citenamefont {Albrecht}\ and\ \citenamefont {Steinhardt}(1982)}]{Albrecht:1982wi}%
  \BibitemOpen
  \bibfield  {author} {\bibinfo {author} {\bibfnamefont {A.}~\bibnamefont {Albrecht}}\ and\ \bibinfo {author} {\bibfnamefont {P.~J.}\ \bibnamefont {Steinhardt}},\ }\href {\doibase 10.1103/PhysRevLett.48.1220} {\bibfield  {journal} {\bibinfo  {journal} {Phys. Rev. Lett.}\ }\textbf {\bibinfo {volume} {48}},\ \bibinfo {pages} {1220} (\bibinfo {year} {1982})}\BibitemShut {NoStop}%
\bibitem [{\citenamefont {Mukhanov}\ and\ \citenamefont {Chibisov}(1981)}]{Mukhanov:1981xt}%
  \BibitemOpen
  \bibfield  {author} {\bibinfo {author} {\bibfnamefont {V.~F.}\ \bibnamefont {Mukhanov}}\ and\ \bibinfo {author} {\bibfnamefont {G.~V.}\ \bibnamefont {Chibisov}},\ }\href@noop {} {\bibfield  {journal} {\bibinfo  {journal} {JETP Lett.}\ }\textbf {\bibinfo {volume} {33}},\ \bibinfo {pages} {532} (\bibinfo {year} {1981})}\BibitemShut {NoStop}%
\bibitem [{\citenamefont {Starobinsky}(1982)}]{Starobinsky:1982ee}%
  \BibitemOpen
  \bibfield  {author} {\bibinfo {author} {\bibfnamefont {A.~A.}\ \bibnamefont {Starobinsky}},\ }\href {\doibase 10.1016/0370-2693(82)90541-X} {\bibfield  {journal} {\bibinfo  {journal} {Phys. Lett. B}\ }\textbf {\bibinfo {volume} {117}},\ \bibinfo {pages} {175} (\bibinfo {year} {1982})}\BibitemShut {NoStop}%
\bibitem [{\citenamefont {Hawking}(1982)}]{Hawking:1982cz}%
  \BibitemOpen
  \bibfield  {author} {\bibinfo {author} {\bibfnamefont {S.~W.}\ \bibnamefont {Hawking}},\ }\href {\doibase 10.1016/0370-2693(82)90373-2} {\bibfield  {journal} {\bibinfo  {journal} {Phys. Lett. B}\ }\textbf {\bibinfo {volume} {115}},\ \bibinfo {pages} {295} (\bibinfo {year} {1982})}\BibitemShut {NoStop}%
\bibitem [{\citenamefont {Guth}\ and\ \citenamefont {Pi}(1982)}]{Guth:1982ec}%
  \BibitemOpen
  \bibfield  {author} {\bibinfo {author} {\bibfnamefont {A.~H.}\ \bibnamefont {Guth}}\ and\ \bibinfo {author} {\bibfnamefont {S.~Y.}\ \bibnamefont {Pi}},\ }\href {\doibase 10.1103/PhysRevLett.49.1110} {\bibfield  {journal} {\bibinfo  {journal} {Phys. Rev. Lett.}\ }\textbf {\bibinfo {volume} {49}},\ \bibinfo {pages} {1110} (\bibinfo {year} {1982})}\BibitemShut {NoStop}%
\bibitem [{\citenamefont {Bardeen}\ \emph {et~al.}(1983)\citenamefont {Bardeen}, \citenamefont {Steinhardt},\ and\ \citenamefont {Turner}}]{Bardeen:1983qw}%
  \BibitemOpen
  \bibfield  {author} {\bibinfo {author} {\bibfnamefont {J.~M.}\ \bibnamefont {Bardeen}}, \bibinfo {author} {\bibfnamefont {P.~J.}\ \bibnamefont {Steinhardt}}, \ and\ \bibinfo {author} {\bibfnamefont {M.~S.}\ \bibnamefont {Turner}},\ }\href {\doibase 10.1103/PhysRevD.28.679} {\bibfield  {journal} {\bibinfo  {journal} {Phys. Rev. D}\ }\textbf {\bibinfo {volume} {28}},\ \bibinfo {pages} {679} (\bibinfo {year} {1983})}\BibitemShut {NoStop}%
\bibitem [{\citenamefont {Weinberg}(2003)}]{Weinberg:2003sw}%
  \BibitemOpen
  \bibfield  {author} {\bibinfo {author} {\bibfnamefont {S.}~\bibnamefont {Weinberg}},\ }\href {\doibase 10.1103/PhysRevD.67.123504} {\bibfield  {journal} {\bibinfo  {journal} {Phys. Rev. D}\ }\textbf {\bibinfo {volume} {67}},\ \bibinfo {pages} {123504} (\bibinfo {year} {2003})},\ \Eprint {http://arxiv.org/abs/astro-ph/0302326} {arXiv:astro-ph/0302326} \BibitemShut {NoStop}%
\bibitem [{\citenamefont {Braglia}\ and\ \citenamefont {Pinol}(2025)}]{Braglia:2025cee}%
  \BibitemOpen
  \bibfield  {author} {\bibinfo {author} {\bibfnamefont {M.}~\bibnamefont {Braglia}}\ and\ \bibinfo {author} {\bibfnamefont {L.}~\bibnamefont {Pinol}},\ }\href@noop {} {\  (\bibinfo {year} {2025})},\ \Eprint {http://arxiv.org/abs/2504.07926} {arXiv:2504.07926 [astro-ph.CO]} \BibitemShut {NoStop}%
\bibitem [{\citenamefont {Ballesteros}\ \emph {et~al.}(2024)\citenamefont {Ballesteros}, \citenamefont {Gamb\'\i{}n~Egea},\ and\ \citenamefont {Riccardi}}]{Ballesteros:2024qqx}%
  \BibitemOpen
  \bibfield  {author} {\bibinfo {author} {\bibfnamefont {G.}~\bibnamefont {Ballesteros}}, \bibinfo {author} {\bibfnamefont {J.}~\bibnamefont {Gamb\'\i{}n~Egea}}, \ and\ \bibinfo {author} {\bibfnamefont {F.}~\bibnamefont {Riccardi}},\ }\href@noop {} {\  (\bibinfo {year} {2024})},\ \Eprint {http://arxiv.org/abs/2411.19674} {arXiv:2411.19674 [hep-th]} \BibitemShut {NoStop}%
\bibitem [{\citenamefont {Choi}\ \emph {et~al.}(2009)\citenamefont {Choi}, \citenamefont {Gong},\ and\ \citenamefont {Jeong}}]{Choi:2008et}%
  \BibitemOpen
  \bibfield  {author} {\bibinfo {author} {\bibfnamefont {K.-Y.}\ \bibnamefont {Choi}}, \bibinfo {author} {\bibfnamefont {J.-O.}\ \bibnamefont {Gong}}, \ and\ \bibinfo {author} {\bibfnamefont {D.}~\bibnamefont {Jeong}},\ }\href {\doibase 10.1088/1475-7516/2009/02/032} {\bibfield  {journal} {\bibinfo  {journal} {JCAP}\ }\textbf {\bibinfo {volume} {02}},\ \bibinfo {pages} {032} (\bibinfo {year} {2009})},\ \Eprint {http://arxiv.org/abs/0810.2299} {arXiv:0810.2299 [hep-ph]} \BibitemShut {NoStop}%
\bibitem [{\citenamefont {Gonz\'alez}\ \emph {et~al.}(2018)\citenamefont {Gonz\'alez}, \citenamefont {Palma},\ and\ \citenamefont {Videla}}]{Gonzalez:2018jax}%
  \BibitemOpen
  \bibfield  {author} {\bibinfo {author} {\bibfnamefont {P.}~\bibnamefont {Gonz\'alez}}, \bibinfo {author} {\bibfnamefont {G.~A.}\ \bibnamefont {Palma}}, \ and\ \bibinfo {author} {\bibfnamefont {N.}~\bibnamefont {Videla}},\ }\href {\doibase 10.1088/1475-7516/2018/12/001} {\bibfield  {journal} {\bibinfo  {journal} {JCAP}\ }\textbf {\bibinfo {volume} {12}},\ \bibinfo {pages} {001} (\bibinfo {year} {2018})},\ \Eprint {http://arxiv.org/abs/1805.10360} {arXiv:1805.10360 [hep-th]} \BibitemShut {NoStop}%
\bibitem [{\citenamefont {Martin}\ and\ \citenamefont {Pinol}(2021)}]{Martin:2021frd}%
  \BibitemOpen
  \bibfield  {author} {\bibinfo {author} {\bibfnamefont {J.}~\bibnamefont {Martin}}\ and\ \bibinfo {author} {\bibfnamefont {L.}~\bibnamefont {Pinol}},\ }\href {\doibase 10.1088/1475-7516/2021/12/022} {\bibfield  {journal} {\bibinfo  {journal} {JCAP}\ }\textbf {\bibinfo {volume} {12}},\ \bibinfo {pages} {022} (\bibinfo {year} {2021})},\ \Eprint {http://arxiv.org/abs/2105.03301} {arXiv:2105.03301 [astro-ph.CO]} \BibitemShut {NoStop}%
\bibitem [{\citenamefont {Auclair}\ and\ \citenamefont {Ringeval}(2022)}]{Auclair:2022yxs}%
  \BibitemOpen
  \bibfield  {author} {\bibinfo {author} {\bibfnamefont {P.}~\bibnamefont {Auclair}}\ and\ \bibinfo {author} {\bibfnamefont {C.}~\bibnamefont {Ringeval}},\ }\href {\doibase 10.1103/PhysRevD.106.063512} {\bibfield  {journal} {\bibinfo  {journal} {Phys. Rev. D}\ }\textbf {\bibinfo {volume} {106}},\ \bibinfo {pages} {063512} (\bibinfo {year} {2022})},\ \Eprint {http://arxiv.org/abs/2205.12608} {arXiv:2205.12608 [astro-ph.CO]} \BibitemShut {NoStop}%
\bibitem [{\citenamefont {Langlois}\ and\ \citenamefont {Vernizzi}(2005)}]{Langlois:2005qp}%
  \BibitemOpen
  \bibfield  {author} {\bibinfo {author} {\bibfnamefont {D.}~\bibnamefont {Langlois}}\ and\ \bibinfo {author} {\bibfnamefont {F.}~\bibnamefont {Vernizzi}},\ }\href {\doibase 10.1103/PhysRevD.72.103501} {\bibfield  {journal} {\bibinfo  {journal} {Phys. Rev. D}\ }\textbf {\bibinfo {volume} {72}},\ \bibinfo {pages} {103501} (\bibinfo {year} {2005})},\ \Eprint {http://arxiv.org/abs/astro-ph/0509078} {arXiv:astro-ph/0509078} \BibitemShut {NoStop}%
\bibitem [{\citenamefont {Maldacena}(2003)}]{Maldacena:2002vr}%
  \BibitemOpen
  \bibfield  {author} {\bibinfo {author} {\bibfnamefont {J.~M.}\ \bibnamefont {Maldacena}},\ }\href {\doibase 10.1088/1126-6708/2003/05/013} {\bibfield  {journal} {\bibinfo  {journal} {JHEP}\ }\textbf {\bibinfo {volume} {05}},\ \bibinfo {pages} {013} (\bibinfo {year} {2003})},\ \Eprint {http://arxiv.org/abs/astro-ph/0210603} {arXiv:astro-ph/0210603} \BibitemShut {NoStop}%
\bibitem [{\citenamefont {Weinberg}(2005)}]{Weinberg:2005vy}%
  \BibitemOpen
  \bibfield  {author} {\bibinfo {author} {\bibfnamefont {S.}~\bibnamefont {Weinberg}},\ }\href {\doibase 10.1103/PhysRevD.72.043514} {\bibfield  {journal} {\bibinfo  {journal} {Phys. Rev. D}\ }\textbf {\bibinfo {volume} {72}},\ \bibinfo {pages} {043514} (\bibinfo {year} {2005})},\ \Eprint {http://arxiv.org/abs/hep-th/0506236} {arXiv:hep-th/0506236} \BibitemShut {NoStop}%
\bibitem [{\citenamefont {Kahya}\ \emph {et~al.}(2011)\citenamefont {Kahya}, \citenamefont {Onemli},\ and\ \citenamefont {Woodard}}]{Kahya:2010xh}%
  \BibitemOpen
  \bibfield  {author} {\bibinfo {author} {\bibfnamefont {E.~O.}\ \bibnamefont {Kahya}}, \bibinfo {author} {\bibfnamefont {V.~K.}\ \bibnamefont {Onemli}}, \ and\ \bibinfo {author} {\bibfnamefont {R.~P.}\ \bibnamefont {Woodard}},\ }\href {\doibase 10.1016/j.physletb.2010.09.050} {\bibfield  {journal} {\bibinfo  {journal} {Phys. Lett. B}\ }\textbf {\bibinfo {volume} {694}},\ \bibinfo {pages} {101} (\bibinfo {year} {2011})},\ \Eprint {http://arxiv.org/abs/1006.3999} {arXiv:1006.3999 [astro-ph.CO]} \BibitemShut {NoStop}%
\bibitem [{\citenamefont {Pimentel}\ \emph {et~al.}(2012)\citenamefont {Pimentel}, \citenamefont {Senatore},\ and\ \citenamefont {Zaldarriaga}}]{Pimentel:2012tw}%
  \BibitemOpen
  \bibfield  {author} {\bibinfo {author} {\bibfnamefont {G.~L.}\ \bibnamefont {Pimentel}}, \bibinfo {author} {\bibfnamefont {L.}~\bibnamefont {Senatore}}, \ and\ \bibinfo {author} {\bibfnamefont {M.}~\bibnamefont {Zaldarriaga}},\ }\href {\doibase 10.1007/JHEP07(2012)166} {\bibfield  {journal} {\bibinfo  {journal} {JHEP}\ }\textbf {\bibinfo {volume} {07}},\ \bibinfo {pages} {166} (\bibinfo {year} {2012})},\ \Eprint {http://arxiv.org/abs/1203.6651} {arXiv:1203.6651 [hep-th]} \BibitemShut {NoStop}%
\bibitem [{\citenamefont {Senatore}\ and\ \citenamefont {Zaldarriaga}(2013{\natexlab{a}})}]{Senatore:2012ya}%
  \BibitemOpen
  \bibfield  {author} {\bibinfo {author} {\bibfnamefont {L.}~\bibnamefont {Senatore}}\ and\ \bibinfo {author} {\bibfnamefont {M.}~\bibnamefont {Zaldarriaga}},\ }\href {\doibase 10.1007/JHEP09(2013)148} {\bibfield  {journal} {\bibinfo  {journal} {JHEP}\ }\textbf {\bibinfo {volume} {09}},\ \bibinfo {pages} {148} (\bibinfo {year} {2013}{\natexlab{a}})},\ \Eprint {http://arxiv.org/abs/1210.6048} {arXiv:1210.6048 [hep-th]} \BibitemShut {NoStop}%
\bibitem [{\citenamefont {Creminelli}\ \emph {et~al.}(2006)\citenamefont {Creminelli}, \citenamefont {Luty}, \citenamefont {Nicolis},\ and\ \citenamefont {Senatore}}]{Creminelli:2006xe}%
  \BibitemOpen
  \bibfield  {author} {\bibinfo {author} {\bibfnamefont {P.}~\bibnamefont {Creminelli}}, \bibinfo {author} {\bibfnamefont {M.~A.}\ \bibnamefont {Luty}}, \bibinfo {author} {\bibfnamefont {A.}~\bibnamefont {Nicolis}}, \ and\ \bibinfo {author} {\bibfnamefont {L.}~\bibnamefont {Senatore}},\ }\href {\doibase 10.1088/1126-6708/2006/12/080} {\bibfield  {journal} {\bibinfo  {journal} {JHEP}\ }\textbf {\bibinfo {volume} {12}},\ \bibinfo {pages} {080} (\bibinfo {year} {2006})},\ \Eprint {http://arxiv.org/abs/hep-th/0606090} {arXiv:hep-th/0606090} \BibitemShut {NoStop}%
\bibitem [{\citenamefont {Cheung}\ \emph {et~al.}(2008)\citenamefont {Cheung}, \citenamefont {Creminelli}, \citenamefont {Fitzpatrick}, \citenamefont {Kaplan},\ and\ \citenamefont {Senatore}}]{Cheung:2007st}%
  \BibitemOpen
  \bibfield  {author} {\bibinfo {author} {\bibfnamefont {C.}~\bibnamefont {Cheung}}, \bibinfo {author} {\bibfnamefont {P.}~\bibnamefont {Creminelli}}, \bibinfo {author} {\bibfnamefont {A.~L.}\ \bibnamefont {Fitzpatrick}}, \bibinfo {author} {\bibfnamefont {J.}~\bibnamefont {Kaplan}}, \ and\ \bibinfo {author} {\bibfnamefont {L.}~\bibnamefont {Senatore}},\ }\href {\doibase 10.1088/1126-6708/2008/03/014} {\bibfield  {journal} {\bibinfo  {journal} {JHEP}\ }\textbf {\bibinfo {volume} {03}},\ \bibinfo {pages} {014} (\bibinfo {year} {2008})},\ \Eprint {http://arxiv.org/abs/0709.0293} {arXiv:0709.0293 [hep-th]} \BibitemShut {NoStop}%
\bibitem [{\citenamefont {Byrnes}\ \emph {et~al.}(2025)\citenamefont {Byrnes}, \citenamefont {Franciolini}, \citenamefont {Harada}, \citenamefont {Pani},\ and\ \citenamefont {Sasaki}}]{Byrnes:2025tji}%
  \BibitemOpen
  \bibinfo {editor} {\bibfnamefont {C.}~\bibnamefont {Byrnes}}, \bibinfo {editor} {\bibfnamefont {G.}~\bibnamefont {Franciolini}}, \bibinfo {editor} {\bibfnamefont {T.}~\bibnamefont {Harada}}, \bibinfo {editor} {\bibfnamefont {P.}~\bibnamefont {Pani}}, \ and\ \bibinfo {editor} {\bibfnamefont {M.}~\bibnamefont {Sasaki}},\ eds.,\ \href@noop {} {\emph {\bibinfo {title} {{Primordial Black Holes}}}},\ Springer Series in Astrophysics and Cosmology\ (\bibinfo  {publisher} {Springer},\ \bibinfo {year} {2025})\BibitemShut {NoStop}%
\bibitem [{\citenamefont {Dom\`enech}(2021)}]{Domenech:2021ztg}%
  \BibitemOpen
  \bibfield  {author} {\bibinfo {author} {\bibfnamefont {G.}~\bibnamefont {Dom\`enech}},\ }\href {\doibase 10.3390/universe7110398} {\bibfield  {journal} {\bibinfo  {journal} {Universe}\ }\textbf {\bibinfo {volume} {7}},\ \bibinfo {pages} {398} (\bibinfo {year} {2021})},\ \Eprint {http://arxiv.org/abs/2109.01398} {arXiv:2109.01398 [gr-qc]} \BibitemShut {NoStop}%
\bibitem [{\citenamefont {Wang}(2014)}]{Wang:2013zva}%
  \BibitemOpen
  \bibfield  {author} {\bibinfo {author} {\bibfnamefont {Y.}~\bibnamefont {Wang}},\ }\href {\doibase 10.1088/0253-6102/62/1/19} {\bibfield  {journal} {\bibinfo  {journal} {Commun. Theor. Phys.}\ }\textbf {\bibinfo {volume} {62}},\ \bibinfo {pages} {109} (\bibinfo {year} {2014})},\ \Eprint {http://arxiv.org/abs/1303.1523} {arXiv:1303.1523 [hep-th]} \BibitemShut {NoStop}%
\bibitem [{\citenamefont {'t~Hooft}(1973)}]{tHooft:1973mfk}%
  \BibitemOpen
  \bibfield  {author} {\bibinfo {author} {\bibfnamefont {G.}~\bibnamefont {'t~Hooft}},\ }\href {\doibase 10.1016/0550-3213(73)90376-3} {\bibfield  {journal} {\bibinfo  {journal} {Nucl. Phys. B}\ }\textbf {\bibinfo {volume} {61}},\ \bibinfo {pages} {455} (\bibinfo {year} {1973})}\BibitemShut {NoStop}%
\bibitem [{\citenamefont {Senatore}\ and\ \citenamefont {Zaldarriaga}(2010)}]{Senatore:2009cf}%
  \BibitemOpen
  \bibfield  {author} {\bibinfo {author} {\bibfnamefont {L.}~\bibnamefont {Senatore}}\ and\ \bibinfo {author} {\bibfnamefont {M.}~\bibnamefont {Zaldarriaga}},\ }\href {\doibase 10.1007/JHEP12(2010)008} {\bibfield  {journal} {\bibinfo  {journal} {JHEP}\ }\textbf {\bibinfo {volume} {12}},\ \bibinfo {pages} {008} (\bibinfo {year} {2010})},\ \Eprint {http://arxiv.org/abs/0912.2734} {arXiv:0912.2734 [hep-th]} \BibitemShut {NoStop}%
\bibitem [{\citenamefont {Xue}\ \emph {et~al.}(2011)\citenamefont {Xue}, \citenamefont {Dasgupta},\ and\ \citenamefont {Brandenberger}}]{Xue:2011hm}%
  \BibitemOpen
  \bibfield  {author} {\bibinfo {author} {\bibfnamefont {W.}~\bibnamefont {Xue}}, \bibinfo {author} {\bibfnamefont {K.}~\bibnamefont {Dasgupta}}, \ and\ \bibinfo {author} {\bibfnamefont {R.}~\bibnamefont {Brandenberger}},\ }\href {\doibase 10.1103/PhysRevD.83.083520} {\bibfield  {journal} {\bibinfo  {journal} {Phys. Rev. D}\ }\textbf {\bibinfo {volume} {83}},\ \bibinfo {pages} {083520} (\bibinfo {year} {2011})},\ \Eprint {http://arxiv.org/abs/1103.0285} {arXiv:1103.0285 [hep-th]} \BibitemShut {NoStop}%
\bibitem [{\citenamefont {Urakawa}\ and\ \citenamefont {Tanaka}(2010)}]{Urakawa:2010it}%
  \BibitemOpen
  \bibfield  {author} {\bibinfo {author} {\bibfnamefont {Y.}~\bibnamefont {Urakawa}}\ and\ \bibinfo {author} {\bibfnamefont {T.}~\bibnamefont {Tanaka}},\ }\href {\doibase 10.1103/PhysRevD.82.121301} {\bibfield  {journal} {\bibinfo  {journal} {Phys. Rev. D}\ }\textbf {\bibinfo {volume} {82}},\ \bibinfo {pages} {121301} (\bibinfo {year} {2010})},\ \Eprint {http://arxiv.org/abs/1007.0468} {arXiv:1007.0468 [hep-th]} \BibitemShut {NoStop}%
\bibitem [{\citenamefont {Urakawa}\ and\ \citenamefont {Tanaka}(2011)}]{Urakawa:2010kr}%
  \BibitemOpen
  \bibfield  {author} {\bibinfo {author} {\bibfnamefont {Y.}~\bibnamefont {Urakawa}}\ and\ \bibinfo {author} {\bibfnamefont {T.}~\bibnamefont {Tanaka}},\ }\href {\doibase 10.1143/PTP.125.1067} {\bibfield  {journal} {\bibinfo  {journal} {Prog. Theor. Phys.}\ }\textbf {\bibinfo {volume} {125}},\ \bibinfo {pages} {1067} (\bibinfo {year} {2011})},\ \Eprint {http://arxiv.org/abs/1009.2947} {arXiv:1009.2947 [hep-th]} \BibitemShut {NoStop}%
\bibitem [{\citenamefont {Senatore}\ and\ \citenamefont {Zaldarriaga}(2013{\natexlab{b}})}]{Senatore:2012nq}%
  \BibitemOpen
  \bibfield  {author} {\bibinfo {author} {\bibfnamefont {L.}~\bibnamefont {Senatore}}\ and\ \bibinfo {author} {\bibfnamefont {M.}~\bibnamefont {Zaldarriaga}},\ }\href {\doibase 10.1007/JHEP01(2013)109} {\bibfield  {journal} {\bibinfo  {journal} {JHEP}\ }\textbf {\bibinfo {volume} {01}},\ \bibinfo {pages} {109} (\bibinfo {year} {2013}{\natexlab{b}})},\ \Eprint {http://arxiv.org/abs/1203.6354} {arXiv:1203.6354 [hep-th]} \BibitemShut {NoStop}%
\bibitem [{\citenamefont {Tanaka}\ and\ \citenamefont {Urakawa}(2011)}]{Tanaka:2011aj}%
  \BibitemOpen
  \bibfield  {author} {\bibinfo {author} {\bibfnamefont {T.}~\bibnamefont {Tanaka}}\ and\ \bibinfo {author} {\bibfnamefont {Y.}~\bibnamefont {Urakawa}},\ }\href {\doibase 10.1088/1475-7516/2011/05/014} {\bibfield  {journal} {\bibinfo  {journal} {JCAP}\ }\textbf {\bibinfo {volume} {05}},\ \bibinfo {pages} {014} (\bibinfo {year} {2011})},\ \Eprint {http://arxiv.org/abs/1103.1251} {arXiv:1103.1251 [astro-ph.CO]} \BibitemShut {NoStop}%
\bibitem [{\citenamefont {Pajer}\ \emph {et~al.}(2013)\citenamefont {Pajer}, \citenamefont {Schmidt},\ and\ \citenamefont {Zaldarriaga}}]{Pajer:2013ana}%
  \BibitemOpen
  \bibfield  {author} {\bibinfo {author} {\bibfnamefont {E.}~\bibnamefont {Pajer}}, \bibinfo {author} {\bibfnamefont {F.}~\bibnamefont {Schmidt}}, \ and\ \bibinfo {author} {\bibfnamefont {M.}~\bibnamefont {Zaldarriaga}},\ }\href {\doibase 10.1103/PhysRevD.88.083502} {\bibfield  {journal} {\bibinfo  {journal} {Phys. Rev. D}\ }\textbf {\bibinfo {volume} {88}},\ \bibinfo {pages} {083502} (\bibinfo {year} {2013})},\ \Eprint {http://arxiv.org/abs/1305.0824} {arXiv:1305.0824 [astro-ph.CO]} \BibitemShut {NoStop}%
\bibitem [{\citenamefont {Seery}(2010)}]{Seery:2010kh}%
  \BibitemOpen
  \bibfield  {author} {\bibinfo {author} {\bibfnamefont {D.}~\bibnamefont {Seery}},\ }\href {\doibase 10.1088/0264-9381/27/12/124005} {\bibfield  {journal} {\bibinfo  {journal} {Class. Quant. Grav.}\ }\textbf {\bibinfo {volume} {27}},\ \bibinfo {pages} {124005} (\bibinfo {year} {2010})},\ \Eprint {http://arxiv.org/abs/1005.1649} {arXiv:1005.1649 [astro-ph.CO]} \BibitemShut {NoStop}%
\bibitem [{\citenamefont {Huenupi}\ \emph {et~al.}(2024)\citenamefont {Huenupi}, \citenamefont {Hughes}, \citenamefont {Palma},\ and\ \citenamefont {Sypsas}}]{Huenupi:2024ksc}%
  \BibitemOpen
  \bibfield  {author} {\bibinfo {author} {\bibfnamefont {J.}~\bibnamefont {Huenupi}}, \bibinfo {author} {\bibfnamefont {E.}~\bibnamefont {Hughes}}, \bibinfo {author} {\bibfnamefont {G.~A.}\ \bibnamefont {Palma}}, \ and\ \bibinfo {author} {\bibfnamefont {S.}~\bibnamefont {Sypsas}},\ }\href {\doibase 10.1103/PhysRevD.110.123536} {\bibfield  {journal} {\bibinfo  {journal} {Phys. Rev. D}\ }\textbf {\bibinfo {volume} {110}},\ \bibinfo {pages} {123536} (\bibinfo {year} {2024})},\ \Eprint {http://arxiv.org/abs/2406.07610} {arXiv:2406.07610 [hep-th]} \BibitemShut {NoStop}%
\bibitem [{\citenamefont {Cheng}\ \emph {et~al.}(2022)\citenamefont {Cheng}, \citenamefont {Lee},\ and\ \citenamefont {Ng}}]{Cheng:2021lif}%
  \BibitemOpen
  \bibfield  {author} {\bibinfo {author} {\bibfnamefont {S.-L.}\ \bibnamefont {Cheng}}, \bibinfo {author} {\bibfnamefont {D.-S.}\ \bibnamefont {Lee}}, \ and\ \bibinfo {author} {\bibfnamefont {K.-W.}\ \bibnamefont {Ng}},\ }\href {\doibase 10.1016/j.physletb.2022.136956} {\bibfield  {journal} {\bibinfo  {journal} {Phys. Lett. B}\ }\textbf {\bibinfo {volume} {827}},\ \bibinfo {pages} {136956} (\bibinfo {year} {2022})},\ \Eprint {http://arxiv.org/abs/2106.09275} {arXiv:2106.09275 [astro-ph.CO]} \BibitemShut {NoStop}%
\bibitem [{\citenamefont {Inomata}\ \emph {et~al.}(2023)\citenamefont {Inomata}, \citenamefont {Braglia}, \citenamefont {Chen},\ and\ \citenamefont {Renaux-Petel}}]{Inomata:2022yte}%
  \BibitemOpen
  \bibfield  {author} {\bibinfo {author} {\bibfnamefont {K.}~\bibnamefont {Inomata}}, \bibinfo {author} {\bibfnamefont {M.}~\bibnamefont {Braglia}}, \bibinfo {author} {\bibfnamefont {X.}~\bibnamefont {Chen}}, \ and\ \bibinfo {author} {\bibfnamefont {S.}~\bibnamefont {Renaux-Petel}},\ }\href {\doibase 10.1088/1475-7516/2023/04/011} {\bibfield  {journal} {\bibinfo  {journal} {JCAP}\ }\textbf {\bibinfo {volume} {04}},\ \bibinfo {pages} {011} (\bibinfo {year} {2023})},\ \bibinfo {note} {[Erratum: JCAP 09, E01 (2023)]},\ \Eprint {http://arxiv.org/abs/2211.02586} {arXiv:2211.02586 [astro-ph.CO]} \BibitemShut {NoStop}%
\bibitem [{\citenamefont {Kristiano}\ and\ \citenamefont {Yokoyama}(2024{\natexlab{a}})}]{Kristiano:2022maq}%
  \BibitemOpen
  \bibfield  {author} {\bibinfo {author} {\bibfnamefont {J.}~\bibnamefont {Kristiano}}\ and\ \bibinfo {author} {\bibfnamefont {J.}~\bibnamefont {Yokoyama}},\ }\href {\doibase 10.1103/PhysRevLett.132.221003} {\bibfield  {journal} {\bibinfo  {journal} {Phys. Rev. Lett.}\ }\textbf {\bibinfo {volume} {132}},\ \bibinfo {pages} {221003} (\bibinfo {year} {2024}{\natexlab{a}})},\ \Eprint {http://arxiv.org/abs/2211.03395} {arXiv:2211.03395 [hep-th]} \BibitemShut {NoStop}%
\bibitem [{\citenamefont {Riotto}(2023)}]{Riotto:2023hoz}%
  \BibitemOpen
  \bibfield  {author} {\bibinfo {author} {\bibfnamefont {A.}~\bibnamefont {Riotto}},\ }\href@noop {} {\  (\bibinfo {year} {2023})},\ \Eprint {http://arxiv.org/abs/2301.00599} {arXiv:2301.00599 [astro-ph.CO]} \BibitemShut {NoStop}%
\bibitem [{\citenamefont {Firouzjahi}(2023)}]{Firouzjahi:2023aum}%
  \BibitemOpen
  \bibfield  {author} {\bibinfo {author} {\bibfnamefont {H.}~\bibnamefont {Firouzjahi}},\ }\href {\doibase 10.1088/1475-7516/2023/10/006} {\bibfield  {journal} {\bibinfo  {journal} {JCAP}\ }\textbf {\bibinfo {volume} {10}},\ \bibinfo {pages} {006} (\bibinfo {year} {2023})},\ \Eprint {http://arxiv.org/abs/2303.12025} {arXiv:2303.12025 [astro-ph.CO]} \BibitemShut {NoStop}%
\bibitem [{\citenamefont {Choudhury}\ \emph {et~al.}(2024)\citenamefont {Choudhury}, \citenamefont {Gangopadhyay},\ and\ \citenamefont {Sami}}]{Choudhury:2023vuj}%
  \BibitemOpen
  \bibfield  {author} {\bibinfo {author} {\bibfnamefont {S.}~\bibnamefont {Choudhury}}, \bibinfo {author} {\bibfnamefont {M.~R.}\ \bibnamefont {Gangopadhyay}}, \ and\ \bibinfo {author} {\bibfnamefont {M.}~\bibnamefont {Sami}},\ }\href {\doibase 10.1140/epjc/s10052-024-13218-2} {\bibfield  {journal} {\bibinfo  {journal} {Eur. Phys. J. C}\ }\textbf {\bibinfo {volume} {84}},\ \bibinfo {pages} {884} (\bibinfo {year} {2024})},\ \Eprint {http://arxiv.org/abs/2301.10000} {arXiv:2301.10000 [astro-ph.CO]} \BibitemShut {NoStop}%
\bibitem [{\citenamefont {Motohashi}\ and\ \citenamefont {Tada}(2023)}]{Motohashi:2023syh}%
  \BibitemOpen
  \bibfield  {author} {\bibinfo {author} {\bibfnamefont {H.}~\bibnamefont {Motohashi}}\ and\ \bibinfo {author} {\bibfnamefont {Y.}~\bibnamefont {Tada}},\ }\href {\doibase 10.1088/1475-7516/2023/08/069} {\bibfield  {journal} {\bibinfo  {journal} {JCAP}\ }\textbf {\bibinfo {volume} {08}},\ \bibinfo {pages} {069} (\bibinfo {year} {2023})},\ \Eprint {http://arxiv.org/abs/2303.16035} {arXiv:2303.16035 [astro-ph.CO]} \BibitemShut {NoStop}%
\bibitem [{\citenamefont {Franciolini}\ \emph {et~al.}(2024)\citenamefont {Franciolini}, \citenamefont {Iovino}, \citenamefont {Taoso},\ and\ \citenamefont {Urbano}}]{Franciolini:2023lgy}%
  \BibitemOpen
  \bibfield  {author} {\bibinfo {author} {\bibfnamefont {G.}~\bibnamefont {Franciolini}}, \bibinfo {author} {\bibfnamefont {A.}~\bibnamefont {Iovino}, \bibfnamefont {Junior.}}, \bibinfo {author} {\bibfnamefont {M.}~\bibnamefont {Taoso}}, \ and\ \bibinfo {author} {\bibfnamefont {A.}~\bibnamefont {Urbano}},\ }\href {\doibase 10.1103/PhysRevD.109.123550} {\bibfield  {journal} {\bibinfo  {journal} {Phys. Rev. D}\ }\textbf {\bibinfo {volume} {109}},\ \bibinfo {pages} {123550} (\bibinfo {year} {2024})},\ \Eprint {http://arxiv.org/abs/2305.03491} {arXiv:2305.03491 [astro-ph.CO]} \BibitemShut {NoStop}%
\bibitem [{\citenamefont {Tasinato}(2023)}]{Tasinato:2023ukp}%
  \BibitemOpen
  \bibfield  {author} {\bibinfo {author} {\bibfnamefont {G.}~\bibnamefont {Tasinato}},\ }\href {\doibase 10.1103/PhysRevD.108.043526} {\bibfield  {journal} {\bibinfo  {journal} {Phys. Rev. D}\ }\textbf {\bibinfo {volume} {108}},\ \bibinfo {pages} {043526} (\bibinfo {year} {2023})},\ \Eprint {http://arxiv.org/abs/2305.11568} {arXiv:2305.11568 [hep-th]} \BibitemShut {NoStop}%
\bibitem [{\citenamefont {Cheng}\ \emph {et~al.}(2024)\citenamefont {Cheng}, \citenamefont {Lee},\ and\ \citenamefont {Ng}}]{Cheng:2023ikq}%
  \BibitemOpen
  \bibfield  {author} {\bibinfo {author} {\bibfnamefont {S.-L.}\ \bibnamefont {Cheng}}, \bibinfo {author} {\bibfnamefont {D.-S.}\ \bibnamefont {Lee}}, \ and\ \bibinfo {author} {\bibfnamefont {K.-W.}\ \bibnamefont {Ng}},\ }\href {\doibase 10.1088/1475-7516/2024/03/008} {\bibfield  {journal} {\bibinfo  {journal} {JCAP}\ }\textbf {\bibinfo {volume} {03}},\ \bibinfo {pages} {008} (\bibinfo {year} {2024})},\ \Eprint {http://arxiv.org/abs/2305.16810} {arXiv:2305.16810 [astro-ph.CO]} \BibitemShut {NoStop}%
\bibitem [{\citenamefont {Fumagalli}(2023)}]{Fumagalli:2023hpa}%
  \BibitemOpen
  \bibfield  {author} {\bibinfo {author} {\bibfnamefont {J.}~\bibnamefont {Fumagalli}},\ }\href@noop {} {\  (\bibinfo {year} {2023})},\ \Eprint {http://arxiv.org/abs/2305.19263} {arXiv:2305.19263 [astro-ph.CO]} \BibitemShut {NoStop}%
\bibitem [{\citenamefont {Maity}\ \emph {et~al.}(2024)\citenamefont {Maity}, \citenamefont {Ragavendra}, \citenamefont {Sethi},\ and\ \citenamefont {Sriramkumar}}]{Maity:2023qzw}%
  \BibitemOpen
  \bibfield  {author} {\bibinfo {author} {\bibfnamefont {S.}~\bibnamefont {Maity}}, \bibinfo {author} {\bibfnamefont {H.~V.}\ \bibnamefont {Ragavendra}}, \bibinfo {author} {\bibfnamefont {S.~K.}\ \bibnamefont {Sethi}}, \ and\ \bibinfo {author} {\bibfnamefont {L.}~\bibnamefont {Sriramkumar}},\ }\href {\doibase 10.1088/1475-7516/2024/05/046} {\bibfield  {journal} {\bibinfo  {journal} {JCAP}\ }\textbf {\bibinfo {volume} {05}},\ \bibinfo {pages} {046} (\bibinfo {year} {2024})},\ \Eprint {http://arxiv.org/abs/2307.13636} {arXiv:2307.13636 [astro-ph.CO]} \BibitemShut {NoStop}%
\bibitem [{\citenamefont {Davies}\ \emph {et~al.}(2024)\citenamefont {Davies}, \citenamefont {Iacconi},\ and\ \citenamefont {Mulryne}}]{Davies:2023hhn}%
  \BibitemOpen
  \bibfield  {author} {\bibinfo {author} {\bibfnamefont {M.~W.}\ \bibnamefont {Davies}}, \bibinfo {author} {\bibfnamefont {L.}~\bibnamefont {Iacconi}}, \ and\ \bibinfo {author} {\bibfnamefont {D.~J.}\ \bibnamefont {Mulryne}},\ }\href {\doibase 10.1088/1475-7516/2024/04/050} {\bibfield  {journal} {\bibinfo  {journal} {JCAP}\ }\textbf {\bibinfo {volume} {04}},\ \bibinfo {pages} {050} (\bibinfo {year} {2024})},\ \Eprint {http://arxiv.org/abs/2312.05694} {arXiv:2312.05694 [astro-ph.CO]} \BibitemShut {NoStop}%
\bibitem [{\citenamefont {Iacconi}\ \emph {et~al.}(2024)\citenamefont {Iacconi}, \citenamefont {Mulryne},\ and\ \citenamefont {Seery}}]{Iacconi:2023ggt}%
  \BibitemOpen
  \bibfield  {author} {\bibinfo {author} {\bibfnamefont {L.}~\bibnamefont {Iacconi}}, \bibinfo {author} {\bibfnamefont {D.}~\bibnamefont {Mulryne}}, \ and\ \bibinfo {author} {\bibfnamefont {D.}~\bibnamefont {Seery}},\ }\href {\doibase 10.1088/1475-7516/2024/06/062} {\bibfield  {journal} {\bibinfo  {journal} {JCAP}\ }\textbf {\bibinfo {volume} {06}},\ \bibinfo {pages} {062} (\bibinfo {year} {2024})},\ \Eprint {http://arxiv.org/abs/2312.12424} {arXiv:2312.12424 [astro-ph.CO]} \BibitemShut {NoStop}%
\bibitem [{\citenamefont {Inomata}(2024)}]{Inomata:2024lud}%
  \BibitemOpen
  \bibfield  {author} {\bibinfo {author} {\bibfnamefont {K.}~\bibnamefont {Inomata}},\ }\href {\doibase 10.1103/PhysRevLett.133.141001} {\bibfield  {journal} {\bibinfo  {journal} {Phys. Rev. Lett.}\ }\textbf {\bibinfo {volume} {133}},\ \bibinfo {pages} {141001} (\bibinfo {year} {2024})},\ \Eprint {http://arxiv.org/abs/2403.04682} {arXiv:2403.04682 [astro-ph.CO]} \BibitemShut {NoStop}%
\bibitem [{\citenamefont {Firouzjahi}(2024{\natexlab{a}})}]{Firouzjahi:2024psd}%
  \BibitemOpen
  \bibfield  {author} {\bibinfo {author} {\bibfnamefont {H.}~\bibnamefont {Firouzjahi}},\ }\href {\doibase 10.1103/PhysRevD.110.043519} {\bibfield  {journal} {\bibinfo  {journal} {Phys. Rev. D}\ }\textbf {\bibinfo {volume} {110}},\ \bibinfo {pages} {043519} (\bibinfo {year} {2024}{\natexlab{a}})},\ \Eprint {http://arxiv.org/abs/2403.03841} {arXiv:2403.03841 [astro-ph.CO]} \BibitemShut {NoStop}%
\bibitem [{\citenamefont {Caravano}\ \emph {et~al.}(2024)\citenamefont {Caravano}, \citenamefont {Inomata},\ and\ \citenamefont {Renaux-Petel}}]{Caravano:2024tlp}%
  \BibitemOpen
  \bibfield  {author} {\bibinfo {author} {\bibfnamefont {A.}~\bibnamefont {Caravano}}, \bibinfo {author} {\bibfnamefont {K.}~\bibnamefont {Inomata}}, \ and\ \bibinfo {author} {\bibfnamefont {S.}~\bibnamefont {Renaux-Petel}},\ }\href {\doibase 10.1103/PhysRevLett.133.151001} {\bibfield  {journal} {\bibinfo  {journal} {Phys. Rev. Lett.}\ }\textbf {\bibinfo {volume} {133}},\ \bibinfo {pages} {151001} (\bibinfo {year} {2024})},\ \Eprint {http://arxiv.org/abs/2403.12811} {arXiv:2403.12811 [astro-ph.CO]} \BibitemShut {NoStop}%
\bibitem [{\citenamefont {Braglia}\ and\ \citenamefont {Pinol}(2024)}]{Braglia:2024zsl}%
  \BibitemOpen
  \bibfield  {author} {\bibinfo {author} {\bibfnamefont {M.}~\bibnamefont {Braglia}}\ and\ \bibinfo {author} {\bibfnamefont {L.}~\bibnamefont {Pinol}},\ }\href {\doibase 10.1007/JHEP08(2024)068} {\bibfield  {journal} {\bibinfo  {journal} {JHEP}\ }\textbf {\bibinfo {volume} {08}},\ \bibinfo {pages} {068} (\bibinfo {year} {2024})},\ \Eprint {http://arxiv.org/abs/2403.14558} {arXiv:2403.14558 [astro-ph.CO]} \BibitemShut {NoStop}%
\bibitem [{\citenamefont {Kawaguchi}\ \emph {et~al.}(2024{\natexlab{a}})\citenamefont {Kawaguchi}, \citenamefont {Tsujikawa},\ and\ \citenamefont {Yamada}}]{Kawaguchi:2024lsw}%
  \BibitemOpen
  \bibfield  {author} {\bibinfo {author} {\bibfnamefont {R.}~\bibnamefont {Kawaguchi}}, \bibinfo {author} {\bibfnamefont {S.}~\bibnamefont {Tsujikawa}}, \ and\ \bibinfo {author} {\bibfnamefont {Y.}~\bibnamefont {Yamada}},\ }\href {\doibase 10.1016/j.physletb.2024.138962} {\bibfield  {journal} {\bibinfo  {journal} {Phys. Lett. B}\ }\textbf {\bibinfo {volume} {856}},\ \bibinfo {pages} {138962} (\bibinfo {year} {2024}{\natexlab{a}})},\ \Eprint {http://arxiv.org/abs/2403.16022} {arXiv:2403.16022 [hep-th]} \BibitemShut {NoStop}%
\bibitem [{\citenamefont {Ballesteros}\ and\ \citenamefont {Egea}(2024)}]{Ballesteros:2024zdp}%
  \BibitemOpen
  \bibfield  {author} {\bibinfo {author} {\bibfnamefont {G.}~\bibnamefont {Ballesteros}}\ and\ \bibinfo {author} {\bibfnamefont {J.~G.}\ \bibnamefont {Egea}},\ }\href {\doibase 10.1088/1475-7516/2024/07/052} {\bibfield  {journal} {\bibinfo  {journal} {JCAP}\ }\textbf {\bibinfo {volume} {07}},\ \bibinfo {pages} {052} (\bibinfo {year} {2024})},\ \Eprint {http://arxiv.org/abs/2404.07196} {arXiv:2404.07196 [astro-ph.CO]} \BibitemShut {NoStop}%
\bibitem [{\citenamefont {Kristiano}\ and\ \citenamefont {Yokoyama}(2024{\natexlab{b}})}]{Kristiano:2024vst}%
  \BibitemOpen
  \bibfield  {author} {\bibinfo {author} {\bibfnamefont {J.}~\bibnamefont {Kristiano}}\ and\ \bibinfo {author} {\bibfnamefont {J.}~\bibnamefont {Yokoyama}},\ }\href {\doibase 10.1088/1475-7516/2024/10/036} {\bibfield  {journal} {\bibinfo  {journal} {JCAP}\ }\textbf {\bibinfo {volume} {10}},\ \bibinfo {pages} {036} (\bibinfo {year} {2024}{\natexlab{b}})},\ \Eprint {http://arxiv.org/abs/2405.12145} {arXiv:2405.12145 [astro-ph.CO]} \BibitemShut {NoStop}%
\bibitem [{\citenamefont {Kawaguchi}\ \emph {et~al.}(2024{\natexlab{b}})\citenamefont {Kawaguchi}, \citenamefont {Tsujikawa},\ and\ \citenamefont {Yamada}}]{Kawaguchi:2024rsv}%
  \BibitemOpen
  \bibfield  {author} {\bibinfo {author} {\bibfnamefont {R.}~\bibnamefont {Kawaguchi}}, \bibinfo {author} {\bibfnamefont {S.}~\bibnamefont {Tsujikawa}}, \ and\ \bibinfo {author} {\bibfnamefont {Y.}~\bibnamefont {Yamada}},\ }\href {\doibase 10.1007/JHEP12(2024)095} {\bibfield  {journal} {\bibinfo  {journal} {JHEP}\ }\textbf {\bibinfo {volume} {12}},\ \bibinfo {pages} {095} (\bibinfo {year} {2024}{\natexlab{b}})},\ \Eprint {http://arxiv.org/abs/2407.19742} {arXiv:2407.19742 [hep-th]} \BibitemShut {NoStop}%
\bibitem [{\citenamefont {Fumagalli}(2025)}]{Fumagalli:2024jzz}%
  \BibitemOpen
  \bibfield  {author} {\bibinfo {author} {\bibfnamefont {J.}~\bibnamefont {Fumagalli}},\ }\href {\doibase 10.1007/JHEP01(2025)108} {\bibfield  {journal} {\bibinfo  {journal} {JHEP}\ }\textbf {\bibinfo {volume} {01}},\ \bibinfo {pages} {108} (\bibinfo {year} {2025})},\ \Eprint {http://arxiv.org/abs/2408.08296} {arXiv:2408.08296 [astro-ph.CO]} \BibitemShut {NoStop}%
\bibitem [{\citenamefont {Caravano}\ \emph {et~al.}(2025)\citenamefont {Caravano}, \citenamefont {Franciolini},\ and\ \citenamefont {Renaux-Petel}}]{Caravano:2024moy}%
  \BibitemOpen
  \bibfield  {author} {\bibinfo {author} {\bibfnamefont {A.}~\bibnamefont {Caravano}}, \bibinfo {author} {\bibfnamefont {G.}~\bibnamefont {Franciolini}}, \ and\ \bibinfo {author} {\bibfnamefont {S.}~\bibnamefont {Renaux-Petel}},\ }\href {\doibase 10.1103/PhysRevD.111.063518} {\bibfield  {journal} {\bibinfo  {journal} {Phys. Rev. D}\ }\textbf {\bibinfo {volume} {111}},\ \bibinfo {pages} {063518} (\bibinfo {year} {2025})},\ \Eprint {http://arxiv.org/abs/2410.23942} {arXiv:2410.23942 [astro-ph.CO]} \BibitemShut {NoStop}%
\bibitem [{\citenamefont {Ruiz}\ and\ \citenamefont {Rey}(2024)}]{Ruiz:2024weh}%
  \BibitemOpen
  \bibfield  {author} {\bibinfo {author} {\bibfnamefont {J.~A.}\ \bibnamefont {Ruiz}}\ and\ \bibinfo {author} {\bibfnamefont {J.}~\bibnamefont {Rey}},\ }\href@noop {} {\  (\bibinfo {year} {2024})},\ \Eprint {http://arxiv.org/abs/2410.09014} {arXiv:2410.09014 [astro-ph.CO]} \BibitemShut {NoStop}%
\bibitem [{\citenamefont {Firouzjahi}(2024{\natexlab{b}})}]{Firouzjahi:2024sce}%
  \BibitemOpen
  \bibfield  {author} {\bibinfo {author} {\bibfnamefont {H.}~\bibnamefont {Firouzjahi}},\ }\href {\doibase 10.3390/universe10120456} {\bibfield  {journal} {\bibinfo  {journal} {Universe}\ }\textbf {\bibinfo {volume} {10}},\ \bibinfo {pages} {456} (\bibinfo {year} {2024}{\natexlab{b}})},\ \Eprint {http://arxiv.org/abs/2411.10253} {arXiv:2411.10253 [hep-ph]} \BibitemShut {NoStop}%
\bibitem [{\citenamefont {Sheikhahmadi}\ and\ \citenamefont {Nassiri-Rad}(2024)}]{Sheikhahmadi:2024peu}%
  \BibitemOpen
  \bibfield  {author} {\bibinfo {author} {\bibfnamefont {H.}~\bibnamefont {Sheikhahmadi}}\ and\ \bibinfo {author} {\bibfnamefont {A.}~\bibnamefont {Nassiri-Rad}},\ }\href@noop {} {\  (\bibinfo {year} {2024})},\ \Eprint {http://arxiv.org/abs/2411.18525} {arXiv:2411.18525 [astro-ph.CO]} \BibitemShut {NoStop}%
\bibitem [{\citenamefont {Inomata}(2025{\natexlab{a}})}]{Inomata:2025bqw}%
  \BibitemOpen
  \bibfield  {author} {\bibinfo {author} {\bibfnamefont {K.}~\bibnamefont {Inomata}},\ }\href@noop {} {\  (\bibinfo {year} {2025}{\natexlab{a}})},\ \Eprint {http://arxiv.org/abs/2502.08707} {arXiv:2502.08707 [astro-ph.CO]} \BibitemShut {NoStop}%
\bibitem [{\citenamefont {Fang}\ \emph {et~al.}(2025)\citenamefont {Fang}, \citenamefont {Lyu}, \citenamefont {Chen},\ and\ \citenamefont {Guo}}]{Fang:2025vhi}%
  \BibitemOpen
  \bibfield  {author} {\bibinfo {author} {\bibfnamefont {C.-J.}\ \bibnamefont {Fang}}, \bibinfo {author} {\bibfnamefont {Z.-H.}\ \bibnamefont {Lyu}}, \bibinfo {author} {\bibfnamefont {C.}~\bibnamefont {Chen}}, \ and\ \bibinfo {author} {\bibfnamefont {Z.-K.}\ \bibnamefont {Guo}},\ }\href@noop {} {\  (\bibinfo {year} {2025})},\ \Eprint {http://arxiv.org/abs/2502.09555} {arXiv:2502.09555 [gr-qc]} \BibitemShut {NoStop}%
\bibitem [{\citenamefont {Firouzjahi}\ and\ \citenamefont {Nikbakht}(2025{\natexlab{a}})}]{Firouzjahi:2025gja}%
  \BibitemOpen
  \bibfield  {author} {\bibinfo {author} {\bibfnamefont {H.}~\bibnamefont {Firouzjahi}}\ and\ \bibinfo {author} {\bibfnamefont {B.}~\bibnamefont {Nikbakht}},\ }\href@noop {} {\  (\bibinfo {year} {2025}{\natexlab{a}})},\ \Eprint {http://arxiv.org/abs/2502.09481} {arXiv:2502.09481 [astro-ph.CO]} \BibitemShut {NoStop}%
\bibitem [{\citenamefont {Firouzjahi}\ and\ \citenamefont {Nikbakht}(2025{\natexlab{b}})}]{Firouzjahi:2025ihn}%
  \BibitemOpen
  \bibfield  {author} {\bibinfo {author} {\bibfnamefont {H.}~\bibnamefont {Firouzjahi}}\ and\ \bibinfo {author} {\bibfnamefont {B.}~\bibnamefont {Nikbakht}},\ }\href@noop {} {\  (\bibinfo {year} {2025}{\natexlab{b}})},\ \Eprint {http://arxiv.org/abs/2502.10287} {arXiv:2502.10287 [astro-ph.CO]} \BibitemShut {NoStop}%
\bibitem [{\citenamefont {Inomata}(2025{\natexlab{b}})}]{Inomata:2025pqa}%
  \BibitemOpen
  \bibfield  {author} {\bibinfo {author} {\bibfnamefont {K.}~\bibnamefont {Inomata}},\ }\href@noop {} {\  (\bibinfo {year} {2025}{\natexlab{b}})},\ \Eprint {http://arxiv.org/abs/2502.12112} {arXiv:2502.12112 [astro-ph.CO]} \BibitemShut {NoStop}%
\bibitem [{\citenamefont {Inomata}\ and\ \citenamefont {Nakama}(2019)}]{Inomata:2018epa}%
  \BibitemOpen
  \bibfield  {author} {\bibinfo {author} {\bibfnamefont {K.}~\bibnamefont {Inomata}}\ and\ \bibinfo {author} {\bibfnamefont {T.}~\bibnamefont {Nakama}},\ }\href {\doibase 10.1103/PhysRevD.99.043511} {\bibfield  {journal} {\bibinfo  {journal} {Phys. Rev. D}\ }\textbf {\bibinfo {volume} {99}},\ \bibinfo {pages} {043511} (\bibinfo {year} {2019})},\ \Eprint {http://arxiv.org/abs/1812.00674} {arXiv:1812.00674 [astro-ph.CO]} \BibitemShut {NoStop}%
\bibitem [{\citenamefont {Bagui}\ \emph {et~al.}(2025)\citenamefont {Bagui} \emph {et~al.}}]{LISACosmologyWorkingGroup:2023njw}%
  \BibitemOpen
  \bibfield  {author} {\bibinfo {author} {\bibfnamefont {E.}~\bibnamefont {Bagui}} \emph {et~al.} (\bibinfo {collaboration} {LISA Cosmology Working Group}),\ }\href {\doibase 10.1007/s41114-024-00053-w} {\bibfield  {journal} {\bibinfo  {journal} {Living Rev. Rel.}\ }\textbf {\bibinfo {volume} {28}},\ \bibinfo {pages} {1} (\bibinfo {year} {2025})},\ \Eprint {http://arxiv.org/abs/2310.19857} {arXiv:2310.19857 [astro-ph.CO]} \BibitemShut {NoStop}%
\bibitem [{\citenamefont {Braglia}\ \emph {et~al.}(2024)\citenamefont {Braglia} \emph {et~al.}}]{LISACosmologyWorkingGroup:2024hsc}%
  \BibitemOpen
  \bibfield  {author} {\bibinfo {author} {\bibfnamefont {M.}~\bibnamefont {Braglia}} \emph {et~al.} (\bibinfo {collaboration} {LISA Cosmology Working Group}),\ }\href {\doibase 10.1088/1475-7516/2024/11/032} {\bibfield  {journal} {\bibinfo  {journal} {JCAP}\ }\textbf {\bibinfo {volume} {11}},\ \bibinfo {pages} {032} (\bibinfo {year} {2024})},\ \Eprint {http://arxiv.org/abs/2407.04356} {arXiv:2407.04356 [astro-ph.CO]} \BibitemShut {NoStop}%
\bibitem [{\citenamefont {Gammal}\ \emph {et~al.}(2025)\citenamefont {Gammal} \emph {et~al.}}]{LISACosmologyWorkingGroup:2025vdz}%
  \BibitemOpen
  \bibfield  {author} {\bibinfo {author} {\bibfnamefont {J.~E.}\ \bibnamefont {Gammal}} \emph {et~al.} (\bibinfo {collaboration} {LISA Cosmology Working Group}),\ }\href {\doibase 10.1088/1475-7516/2025/05/062} {\bibfield  {journal} {\bibinfo  {journal} {JCAP}\ }\textbf {\bibinfo {volume} {05}},\ \bibinfo {pages} {062} (\bibinfo {year} {2025})},\ \Eprint {http://arxiv.org/abs/2501.11320} {arXiv:2501.11320 [astro-ph.CO]} \BibitemShut {NoStop}%
\bibitem [{\citenamefont {Cecchini}\ \emph {et~al.}(2025)\citenamefont {Cecchini}, \citenamefont {Franciolini},\ and\ \citenamefont {Pieroni}}]{Cecchini:2025oks}%
  \BibitemOpen
  \bibfield  {author} {\bibinfo {author} {\bibfnamefont {C.}~\bibnamefont {Cecchini}}, \bibinfo {author} {\bibfnamefont {G.}~\bibnamefont {Franciolini}}, \ and\ \bibinfo {author} {\bibfnamefont {M.}~\bibnamefont {Pieroni}},\ }\href {\doibase 10.1103/nxx5-gx7d} {\bibfield  {journal} {\bibinfo  {journal} {Phys. Rev. D}\ }\textbf {\bibinfo {volume} {111}},\ \bibinfo {pages} {123536} (\bibinfo {year} {2025})},\ \Eprint {http://arxiv.org/abs/2503.10805} {arXiv:2503.10805 [astro-ph.CO]} \BibitemShut {NoStop}%
\end{thebibliography}%

\end{document}